\documentclass[trackchanges]{aastex701}

\usepackage{float}
\usepackage{subcaption}

\begin{document}

\title{MAS SKY: On-sky demonstration of the MAS-CCD capabilities}

\author[]{Agustin J. Lapi}
\affiliation{Department of Astronomy and Astrophysics, University of Chicago, Chicago, IL, USA.}
\affiliation{Fermi National Accelerator Laboratory, Batavia, IL, USA.}
\affiliation{Instituto de Inv. en Ing. Eléctrica (IIIE), DIEC, Universidad Nacional del Sur (UNS)-CONICET, Bahía Blanca, Argentina}
\email{ajlapi@uchicago.edu}

\author[]{Miqueas E. Gamero}
\affiliation{Instituto de Inv. en Ing. Eléctrica (IIIE), DIEC, Universidad Nacional del Sur (UNS)-CONICET, Bahía Blanca, Argentina}
\affiliation{Fermi National Accelerator Laboratory, Batavia, IL, USA.}
\email{miqueas.gamero@uns.edu.ar}

\author[]{Christopher Montalban Muñoz}
\affiliation{Departamento Astronomía, Universidad de La Serena. Av. Raúl Bitran 1305, La Serena, Región de Coquimbo, Chile}
\email{}

\author[]{Blas J. Irigoyen Gimenez}
\affiliation{Instituto de Inv. en Ing. Eléctrica (IIIE), DIEC, Universidad Nacional del Sur (UNS)-CONICET, Bahía Blanca, Argentina}
\affiliation{Fermi National Accelerator Laboratory, Batavia, IL, USA.}
\email{}

\author[]{Marco Bonatti}
\affiliation{NSF National Optical-Infrared Astronomy Research Laboratory (NOIRLab)}
\email{}

\author[]{Guillermo Fernández Moroni}
\affiliation{Fermi National Accelerator Laboratory, Batavia, IL, USA.}
\affiliation{Department of Astronomy and Astrophysics, University of Chicago, Chicago, IL, USA.}
\email{gfmoroni@fnal.gov}

\author[]{Braulio Cancino}
\affiliation{NSF National Optical-Infrared Astronomy Research Laboratory (NOIRLab)}
\email{}

\author[]{Fernando Chierchie}
\affiliation{Instituto de Inv. en Ing. Eléctrica (IIIE), DIEC, Universidad Nacional del Sur (UNS)-CONICET, Bahía Blanca, Argentina}
\email{}

\author[]{Gustavo Cancelo}
\affiliation{Fermi National Accelerator Laboratory, Batavia, IL, USA.}
\email{}

\author[]{Claudio Rodrigo Chavez Blanco}
\affiliation{Fermi National Accelerator Laboratory, Batavia, IL, USA.}
\email{}

\author[]{Guillermo Damke}
\affiliation{NSF National Optical-Infrared Astronomy Research Laboratory (NOIRLab)} \email{guillermo.damke@noirlab.edu}

\author[]{Alex Drlica-Wagner}
\affiliation{Fermi National Accelerator Laboratory, Batavia, IL, USA.}
\affiliation{Department of Astronomy and Astrophysics, University of Chicago, Chicago, IL, USA.}
\email{}

\author[]{Juan Estrada}
\affiliation{Brookhaven National Laboratory, Upton, NY, USA.}
\affiliation{Fermi National Accelerator Laboratory, Batavia, IL, USA.}
\affiliation{Department of Astronomy and Astrophysics, University of Chicago, Chicago, IL, USA.}
\email{}

\author[]{Rodrigo Hinojosa}
\affiliation{NSF National Optical-Infrared Astronomy Research Laboratory (NOIRLab)}
\email{rodrigo.hinojosa@noirlab.edu}

\author[]{Stephen Holland}
\affiliation{Lawrence Berkeley National Laboratory, Berkeley, CA, USA.}
\email{}

\author[]{Marcelo Jaque Arancibia}
\affiliation{Departamento Astronomía, Universidad de La Serena. Av. Raúl Bitran 1305, La Serena, Región de Coquimbo, Chile}
\email{marcelo.jaque@userena.cl}

\author[]{José Luis Nilo Castellon}
\affiliation{Departamento Astronomía, Universidad de La Serena. Av. Raúl Bitran 1305, La Serena, Región de Coquimbo, Chile}
\email{jnilo@userena.cl}

\author[]{Javier Tiffenberg}
\affiliation{Fermi National Accelerator Laboratory, Batavia, IL, USA.}
\affiliation{Universidad de Buenos Aires, CABA, Argentina.}
\email{}

\author[]{Sho Uemura}
\affiliation{Fermi National Accelerator Laboratory, Batavia, IL, USA.}
\email{}

\newcommand{\meg}[1]{\textcolor{black}{#1}}
\newcommand{\bj}[1]{\textcolor{black}{#1}}
\newcommand{\al}[1]{\textcolor{black}{#1}}
\newcommand{\ajl}[1]{\textcolor{magenta}{#1}}
\newcommand{\gfm}[1]{\textcolor{blue}{#1}}

\begin{abstract}

The Multiple Amplifier Sensing CCD (MAS-CCD) is a low-noise silicon imager designed to reduce the readout time required, when compared to conventional Skipper-CCDs, to reach the single-photon counting regime by distributing repeated non-destructive charge measurements across multiple skipper output stages. This work presents the first on-sky characterization of a 16-amplifier MAS-CCD installed on the SMARTS-GSU 0.9-meter telescope at the Cerro Tololo Inter-American Observatory (CTIO). The sensor was operated on-sky during a two-week campaign, allowing its performance to be evaluated in a real observatory environment. We report measurements of readout noise, node removal inefficiency associated with the multiple amplifier structure, and noise standard deviation reduction when using multiple samples. Particular attention is given to the validation of previously developed optimization strategies, including mitigation of node removal inefficiency and readout modes aimed at reducing unnecessary overhead. The results demonstrate that the MAS-CCD architecture can preserve its low-noise performance under telescope operating conditions while enabling faster readout than conventional Skipper-CCD operation. These measurements support the suitability of MAS-CCD technology for future astronomical instruments requiring large-area, low-noise imaging with reduced readout time, when compared to conventional Skipper-CCDs.

\end{abstract}

\keywords{Multiple Amplifier Sensing CCD, MAS-CCD, Skipper-CCD, low-noise silicon imagers, non-destructive readout, single-photon counting, astronomical detectors, on-sky characterization, fast readout, node removal inefficiency.}

\section{Introduction} \label{sec:Intro}
The next generation of astronomical instruments will require detectors capable of measuring extremely faint signals. For example, the Stage-5 Spectroscopic Facility (Spec-S5) is a proposed ground-based instrument designed to conduct an extensive spectroscopic survey of distant galaxies for studies of cosmic expansion \citep{besuner2025spectroscopic}. Similarly, the Habitable Worlds Observatory (HWO), which aims to observe biosignatures in the atmospheres of potentially habitable exoplanets \citep{Clery:2023}, will require photon-counting detectors operating at near-ultraviolet, visible, and near-infrared wavelengths to enable direct imaging and spectroscopy of Earth-like planets orbiting Sun-like stars \citep{2024JATIS..10c4006S,2026arXiv260111803F}. These science cases require the detection of extremely faint signals, often at rates less than a single photon per pixel per minute \citep{RauscherNASA2019}.
In this single-photon regime, the Skipper charge-coupled device (Skipper-CCD) has emerged as a particularly relevant detector technology.

This device incorporates a floating-gate architecture in its output stage enabling non-destructive charge readout, allowing multiple measurements of the same charge packet. As a result, the readout noise can be reduced to the single-photon counting regime by averaging repeated samples. The standard deviation of the noise follows a reduction of the form
\begin{equation}
    \sigma = \frac{\sigma_0}{\sqrt{N_S}},
\end{equation}
where $\sigma_0$ is the single-sample readout noise standard deviation and $N_S$ is the number of samples per pixel. For more details on Skipper-CCD operation, see \citet{janesick2001scientific, skipper_2012, Tiffenberg:2017aac}. However, some astronomical applications require reaching this photon-counting regime within a readout time of only a few tens of seconds to readout the entire sensor. To improve the readout time, the Multiple Amplifier Sensing CCD (MAS-CCD) features multiple Skipper readout stages distributed along its serial register. This design enables single-photon counting performance on a shorter timescale than state-of-the-art Skipper-CCD devices \citep{Holland_SDW23, Blas2024}. The extent of the readout-time reduction depends on the number of amplifiers incorporated in the device.

\begin{figure}[h!]
    \centering
    \includegraphics[width=0.95\linewidth]{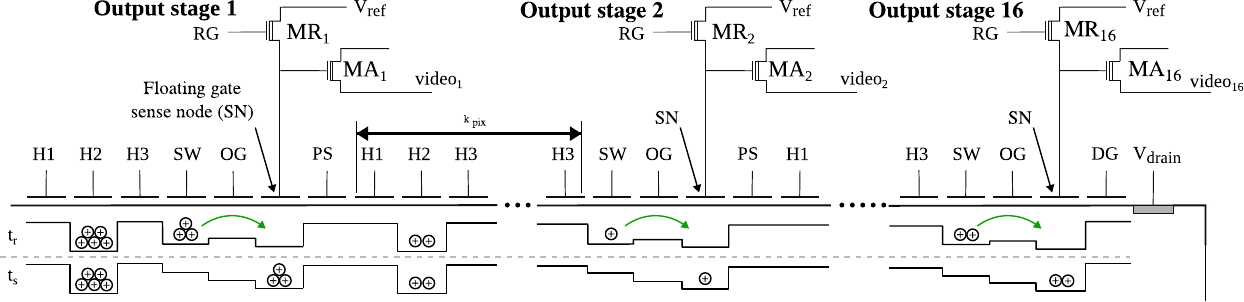}
    \caption{MAS-CCD output stage diagram featuring 16 readout stages. The readout stages are separated by a fixed number of pixels, $k_{pix}$. A qualitative potential diagram of the channel illustrates the charge transfer mechanism across the multiple stages.}
    \label{fig:mas-schematic}
\end{figure}

The MAS-CCD utilized throughout this article features 16 serial Skipper stages. A diagram illustrating its readout architecture is presented in Figure~\ref{fig:mas-schematic}. During readout, charge packets are transported sequentially through the pixels between adjacent amplifiers using the three-phase clocking scheme defined by the H1, H2, and H3 gates. Prior to signal measurement, the charge packets being processed by the different output stages are positioned beneath their respective summing well (SW) gates simultaneously. This condition is depicted as the instant $t_{r}$ in the schematic, when the output stage records the reference level. In the next step, at instant $t_{s}$, each charge packet is transferred beneath the respective sense node (SN), where the signal level is measured simultaneously in all stages. The charge signal is then obtained using the digital Dual Slope Integrator (DSI) technique, i.e., by subtracting the pedestal level from the signal level. As in Skipper-CCDs, this process can be repeated an arbitrary number of times. Once the measurement of the charge packet is complete, it is reinserted into the serial register through the pixel separation (PS) gate and transferred toward the next readout stage, where the process is repeated. \al{These operations are performed in parallel across all output stages.} After the charge packet has been read out by the last serial amplifier, it is dumped to the ohmic contact $\text{V}_{\text{drain}}$ using the dump gate (DG).

The expected readout noise, therefore, depends on both the number of amplifiers and the samples acquired per amplifier. It is expressed as
\begin{equation}
    \sigma = \frac{\sigma_0}{\sqrt{N_S}\sqrt{N_a}},
    \label{ec:noise_reduction}
\end{equation}
where $N_a$ is the number of amplifiers. This reduction is obtained by combining the images produced by each amplifier offline. Each of the $N_a$ images contains the same real charge per pixel but have different realizations of the readout noise sampled by each amplifier.

Basic operating principles for an eight-amplifier MAS-CCD were first demonstrated in \citet{Blas2024}, including lossless charge transfer between independent floating-gate amplifiers, noise reduction through the combination of multiple amplifier measurements, and single-electron resolution using multiple non-destructive samples per amplifier. This work established the feasibility of using multiple inline amplifiers to achieve quantum-level readout faster than a standard Skipper-CCD.

Subsequent studies presented the operation and characterization of a front-illuminated 16-amplifier MAS-CCD aimed at future astronomical instruments \citep{Lapi2024MAS16, Lin_2024}. In particular, \citep{Lapi2024MAS16} demonstrated approximately 1 e$^-$ rms noise with one sample per amplifier, subelectron noise with multiple samples, gain equalization, optimal weighted averaging, common-noise decorrelation, readout speeds of 80kpix per second, and a framework for modeling and characterizing node removal inefficiency.

The process of moving charge through multiple sense nodes on the serial register introduces an additional performance parameter specific to the MAS-CCD: the node removal inefficiency (NRI). NRI quantifies the charge-transfer inefficiency associated with removing the charge packet from the sense node and transferring it back to the subsequent pixels of the serial register.
The potential gradient between these structures, transfer time, and pixel charge occupancy are the main contributing factors for the NRI \citep{Lapi2024MAS16}. A method to reduce NRI was proposed and verified in \citet{irigoyen2025multiple}. The method consists of clocking the $\text{V}_{\text{ref}}$ signal, which is normally held at a fixed reference value, to facilitate a potential gradient toward the H1 gate.

Several time-optimized MAS-CCD readout modes were explored using both 8- and 16-amplifier devices \citep{Blas_SPIE_2024}. Among them, a truncated prescan mode (called ``continuous readout mode'' by \cite{Blas_SPIE_2024}) to avoid the overhead time from the readout of prescan pixels in the extended serial register, and a region-of-interest readout mode, in which additional non-destructive samples are concentrated only in selected image regions. Both approaches reduce readout overhead while preserving the MAS-CCD noise advantage.

In this article, we assess the suitability of different MAS-CCD operating modes for astronomical observations. To evaluate the sensor under real observing conditions, we installed the MAS-CCD on the 0.9-m SMARTS-GSU telescope at the Cerro Tololo Inter-American Observatory (CTIO) in Chile. The primary goal of this campaign was to demonstrate the advantages of the MAS-CCD in improving the signal-to-noise ratio of faint astronomical signals when the operational parameters of the device are optimized and deployed in the field.

The paper is organized as follows. The next section covers the characterization and optimization of the sensor operation in the laboratory. Section \ref{sec:camera_integration} details the camera instrumentation and telescope integration work. Section \ref{sec:engineering_on_field_characterization} revisits the performance and characterization of the sensor during on-sky observations. We conclude in Section \ref{sec:conclusions} with the final remarks.

\section{Sensor characterization before commissioning}
The devices used for this work consists of $N_a = 16$ non-destructive skipper readout stages placed along an extended serial register separated by $k_{pix} = 15$ pixels. The imaging area of the detector consists of a $1024\times512$ array of $15 \mu \text{m}\times15 \mu \text{m}$ pixels. They are $p$-channel, $250\mu \text{m}$-thick MAS-CCDs fabricated on high-resistivity $n$-type silicon ($\sim 10 \text{k}\Omega$cm). The sensors were designed at LBNL to operate as thick, fully depleted devices with high quantum efficiency over a broad wavelength range \citep{Holland:2003} (see \citealt{Holland_SDW23} for more details on the fabricated sensors). They were fabricated under the auspices of the U.S. Department Of Energy (DOE) Quantum Information Science initiative at Teledyne DALSA, diced at LBNL, and packaged and tested at Fermilab. The sensors followed a hybrid fabrication model used for the DECam and DESI CCDs and were back-processed at Microsystem Labs \citep{holland_back_illuminated}, where specialized processing of the back surface, metallization, and anti-reflective coating were applied to increase their collection efficiency.

\begin{figure}[b]
\centering
\begin{subfigure}[t]{0.48\linewidth}
    \centering
    \includegraphics[trim={0 3.2cm 0 2.5cm}, clip=true, width=\linewidth]{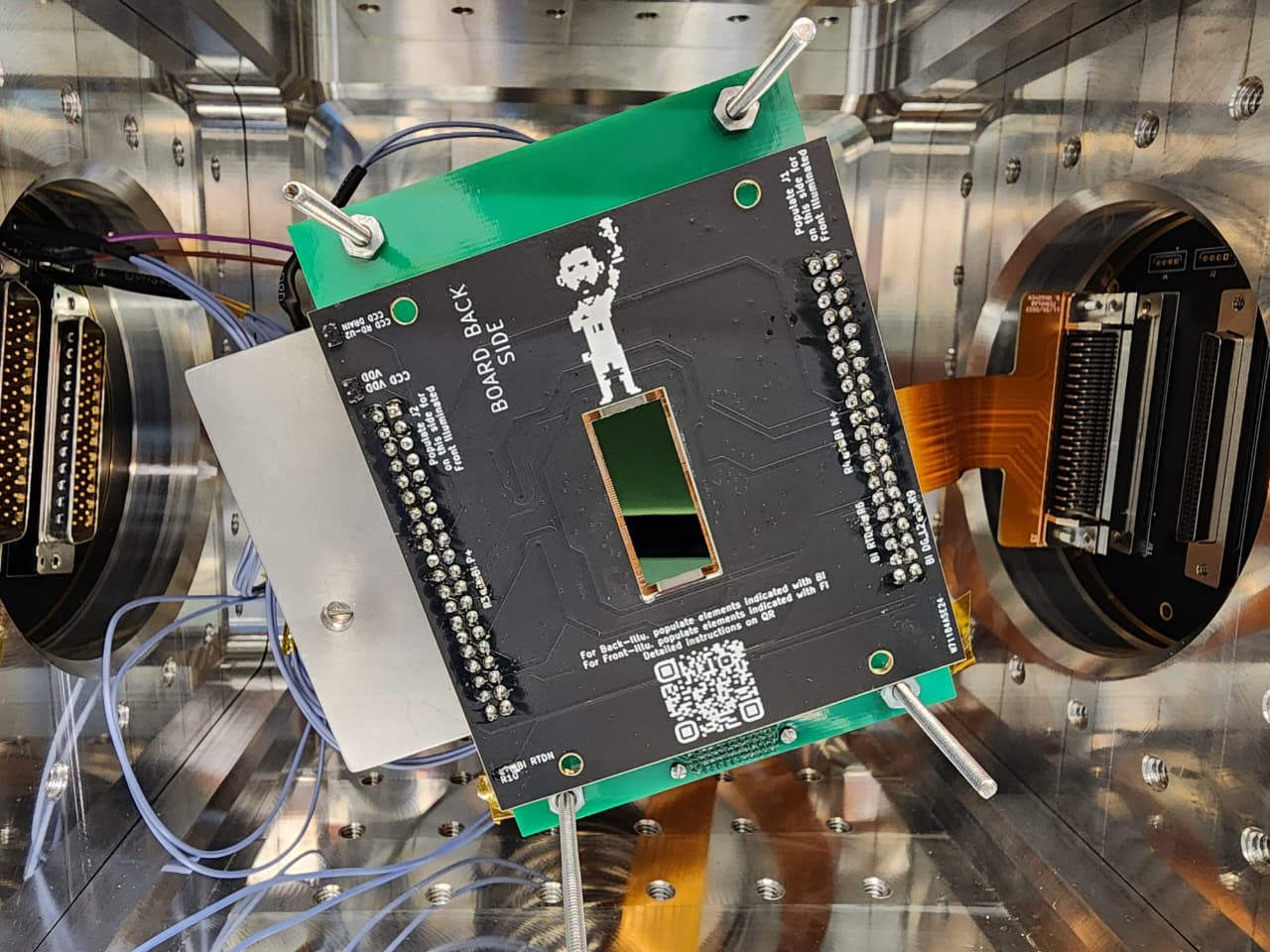}
    \caption{}
    \label{fig:sensor_at_fnal_a}
\end{subfigure}
\hfill
\begin{subfigure}[t]{0.48\linewidth}
    \centering
    \includegraphics[trim={2.5cm 0 0 0}, clip=true, width=\linewidth]{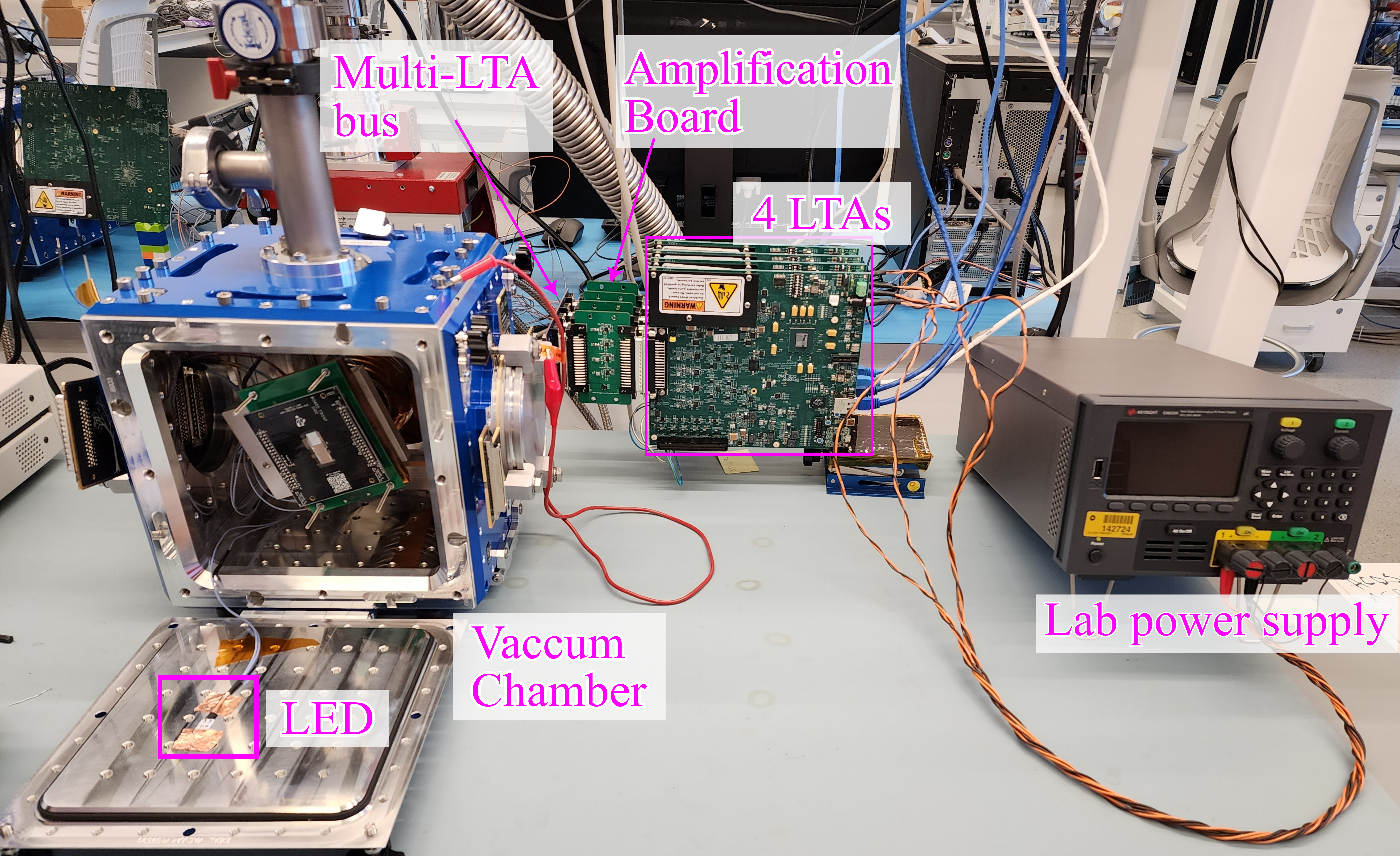}
    \caption{}
    \label{fig:sensor_at_fnal_b}
\end{subfigure}

\caption{(a) Back-illuminated 16-channel MAS-CCD in the experimental setup for characterization purposes at Fermilab. (b) Complete readout electronics system, where the multi-LTA bus, the 4 LTA boards, and the laboratory power supply are indicated.}
\label{fig:sensor_at_fnal}

\end{figure}

The sensors were packaged in a picture-frame-style mechanical package and characterized at Fermilab. The package, depicted in Figure \ref{fig:sensor_at_fnal_a}, is operated inside a vacuum chamber and mechanically attached to a cold finger. A JFET board drives the video signals through a flex cable that connects to a vacuum interface board, which provides the electrical feedthrough between the electronics inside and outside the vacuum chamber. The complete experimental setup is shown in Figure~\ref{fig:sensor_at_fnal_b}. This testing station includes a light-emitting diode (LED) placed at the front center of the sensor to perform time-controlled exposures for full well capacity (FW) and NRI characterization.

The readout system includes four Low Threshold Acquisition (LTA) boards \citep{cancelo2021low}, three of which are configured in follower mode and the remaining one in leader mode. Each LTA board handles four channels; therefore, four boards are required for the sixteen-channel sensor. Following the signal path outside the chamber, a multi-LTA bus distributes the 16 output channels to the controller boards. The video signals then pass through pre-amplification boards to increase their amplitude before entering the controller electronics. The system is powered by a laboratory power supply. The LTA digitizes the signals and performs digital correlated double sampling to compute the pixel values. A Polycold cryocooler, a resistive heater, and a temperature controller allow the sensor to operate at 140 K during the characterization measurements at Fermilab.

Two candidate sensors were characterized and evaluated for their use in the astronomical run at the CTIO. All sixteen amplifiers functioned for both sensors, and no cosmetic defects were identified. This article presents the results for the sensor that was used during the observation run at CTIO. The key parameters evaluated are gain, readout noise, spurious charge (SC), horizontal charge transfer inefficiency (CTI), FW, and node removal inefficiency (NRI). The readout speed was set to 70 kpix/s as used in previous studies of the sensor performance \citep{Lapi2024MAS16}. This results in a single-sample image readout time of $\sim16.7$ seconds.

An optimization procedure was performed to determine the clock voltages that provide the best trade-off between FW and horizontal SC. The selected clock voltages had to support charge packets of at least $\sim50{,}000$ electrons to satisfy the minimum requirements of astronomical instruments, while maintaining the SC below $1.95\times10^{-3}$e$^-$/pix/transfer. This limit was chosen so that the SC contribution to the pixel uncertainty remains comparable to that introduced by the readout noise after averaging the signals from the 16 amplifiers in a single-sample readout. The following subsections present the final laboratory characterization for the main parameters of interest of the selected device for the run.

\subsection{Readout noise performance}
\label{sec:noiseatFermi}
\begin{figure}[t!]
\centering
\begin{subfigure}[t]{0.32\linewidth}
    \centering
    \includegraphics[trim={0 0 0 0}, clip=true, width=\linewidth]{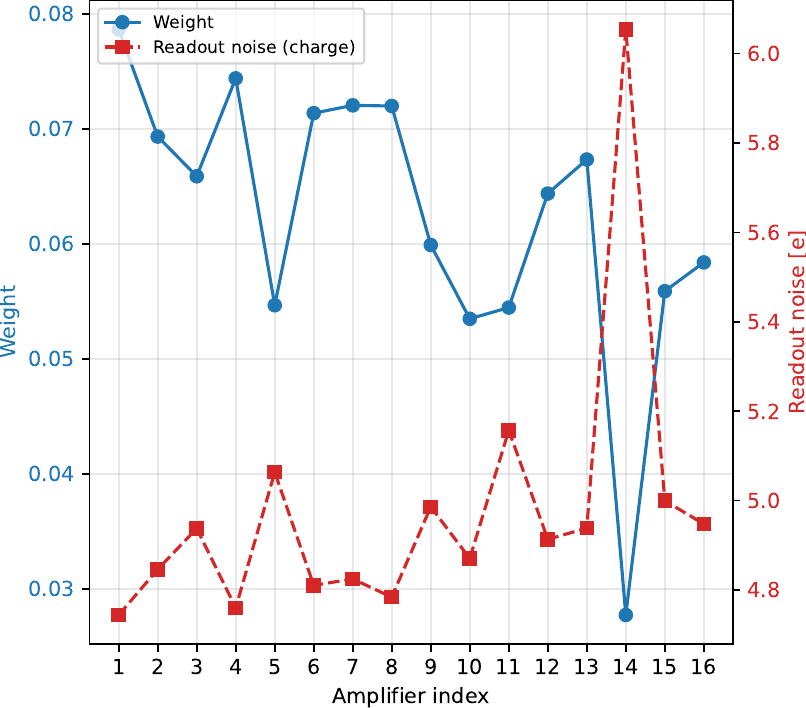}
    \caption{}
    \label{fig:atFermi_a}
\end{subfigure}
\hfill
\begin{subfigure}[t]{0.32\linewidth}
    \centering
    \includegraphics[trim={0 0 0 0}, clip=true, width=\linewidth]{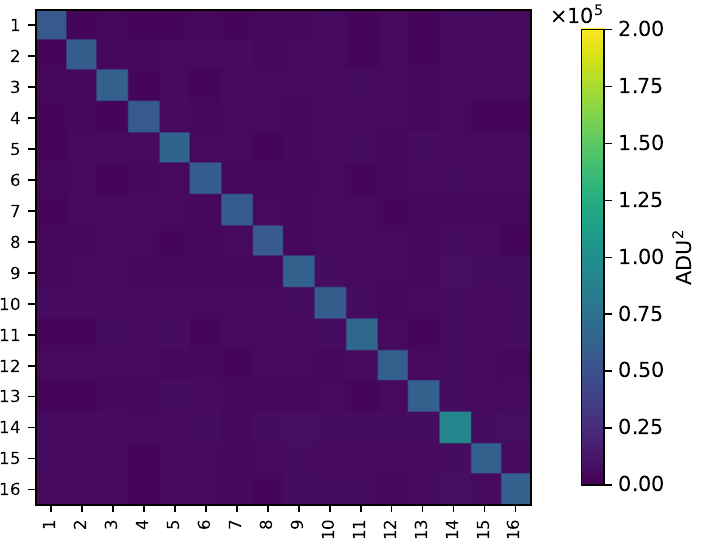}
    \caption{}
    \label{fig:atFermi_b}
\end{subfigure}
\hfill
\begin{subfigure}[t]{0.32\linewidth}
    \centering
    \includegraphics[trim={0 0 0 0}, clip=true, width=\linewidth]{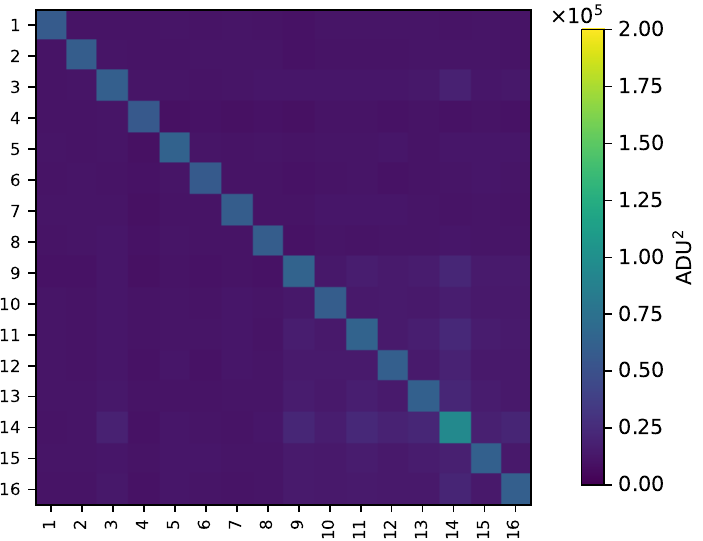}
    \caption{}
    \label{fig:atFermi_c}
\end{subfigure}

\caption{Summary of performance for one sample per pixel ($N_S = 1$) at Fermilab: (a) Standard deviation of the readout noise (red squares) and the optimal weights (blue dots) to average the information from the different channels; (b) correlation matrix of the readout noise among different channels for pixels in charge phase, and (c) similar matrix for the pixels in time phase. }
\label{fig:atFermi}

\end{figure}

This subsection presents standard Skipper-CCD metrics, such as readout noise as a function of the independent samples, $N_S$, as well as other performance measurements intrinsic to MAS-CCDs architecture, such as the optimal combination of output images. Figure \ref{fig:atFermi} presents a summary of the noise performance of the sensor at the Fermilab CCD Lab. The curve on the red axis in Fig.~\ref{fig:atFermi_a} shows the readout noise for each amplifier as measured from the standard deviation of the serial overscan pixels. With the exception of the 14th amplifier, the noise hovers in the range $4.8\,$e$^-$ to $5.2\,$e$^-$ and increases slowly toward the last amplifier. This apparent increase in readout noise may be attributed to the accumulation of spurious charge in the overscan pixels as they are transferred through the extended serial register \citep{gamero2026spurius}.
To compose the mixed image, the optimal weights shown on the blue axis are utilized. These were calculated as introduced in \citet{Lapi2024MAS16} and allow the final image to attain a readout noise as low as $1.16\,$e$^-$ at a single sample per pixel.

To provide insight into common noise sources affecting more than one amplifier, Figures \ref{fig:atFermi_b} and \ref{fig:atFermi_c} show two cross-channel noise covariance matrices for two different time offsets. When the pixels are aligned with the expected charge signal (charge phase), the individual amplifier images are charge-aligned for mixing into the final image. For consecutive amplifiers, this is a shift of $k_{\text{pix}}$ (or multiples of $k_{\text{pix}}$) pixels in the output images, where $k_{\text{pix}}$ is the number of pixels between amplifiers as shown in Figure \ref{fig:mas-schematic}. In this case, the off-diagonal elements exhibit little correlation between amplifiers, which is desirable to prevent very low-frequency noise and charge noise sources during readout.

\begin{table}[t!]
\centering
    \begin{tabular}{|c|c|c|c|c|}
        \hline
        \textbf{LTA number} & 1 & 2 & 3 & 4 \\
        \textbf{Configuration} &  Leader & Follower & Follower & Follower \\
        \textbf{Amplifier index} & 8, 2, 1, 7 & 6, 4, 3, 5 & 14, 12, 13, 11 & 16, 9, 10, 15 \\
        \hline
    \end{tabular}
    \caption{Routing of MAS-CCD amplifier outputs to LTA channels at Fermilab.}
    \label{tab:routing_fnal}
\end{table}

The matrix in Figure \ref{fig:atFermi_c} accounts for correlations in time phase, the cross-covariance is calculated from pixels read simultaneously by the different amplifiers, i.e., no shift is applied to the output images during its calculation. For the ninth amplifier and beyond, the covariance is slightly higher, corresponding to the noisier channels observed in the noise plot in Figure \ref{fig:atFermi_a}. Following the physical routing of the MAS-CCD channels into the LTA controllers indicated in Table \ref{tab:routing_fnal}, the larger correlation appears between the channel outputs of controllers 3 and 4, this effect is unexpected and could be due to grounding.

\begin{figure}[h!]
    \centering
    \includegraphics[width=0.65\linewidth]{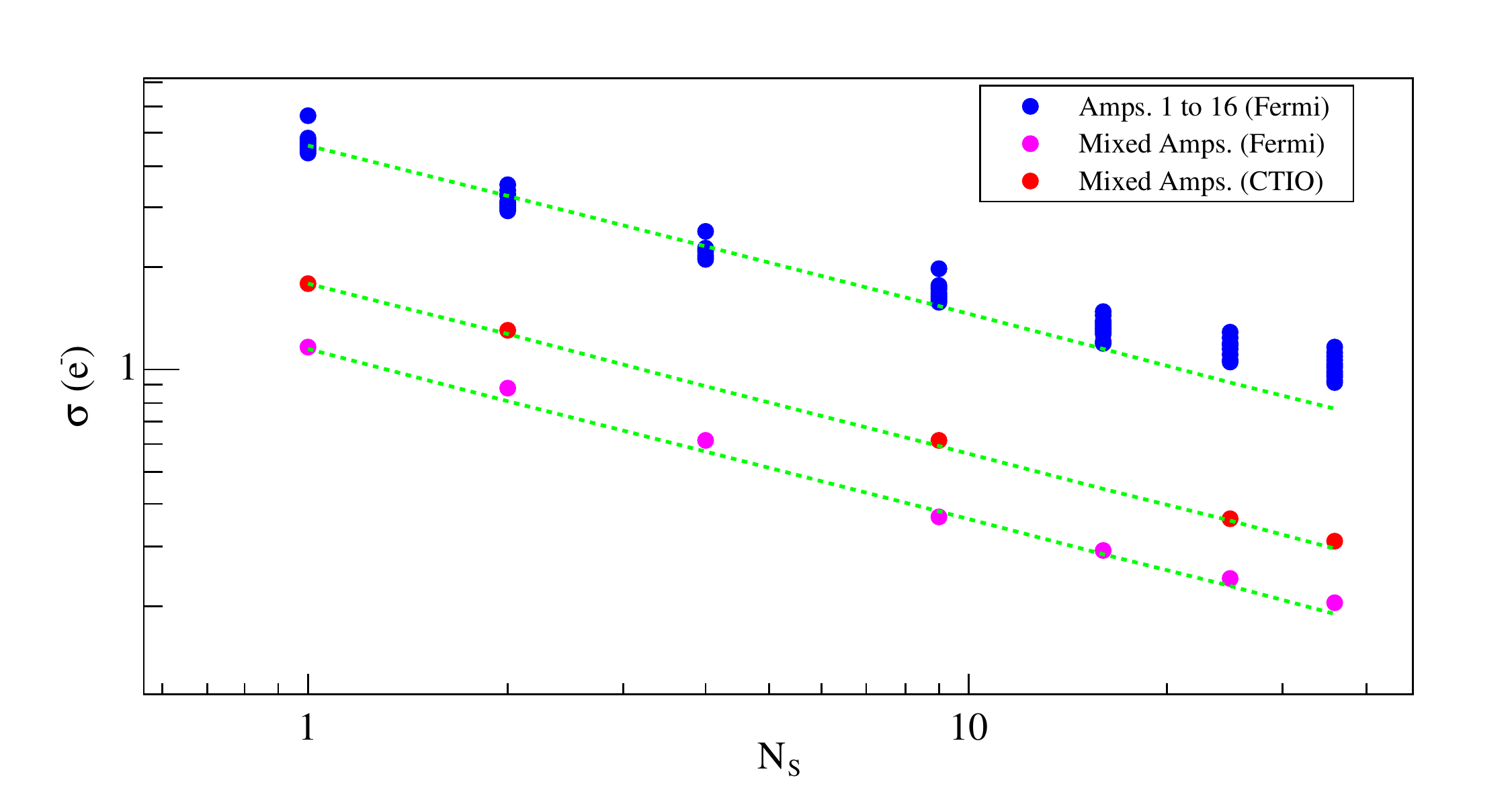}
    \caption{Readout noise estimated from the standard deviation of serial overscan pixels, after combining different numbers of samples, $N_S$. Measurements from Fermilab are shown for each individual amplifier (blue dots) and the optimally mixed image (magenta dots). The red dots indicate the standard readout performance of the mixed channels at CTIO. A theoretical line following the trend $\propto 1/\sqrt{N_S}$ starting from the measured value for $N_S=1$ is also plotted for each cases.}
    \label{fig:noiseAtFermi}
\end{figure}

Having characterized the channel-dependent noise and its correlations, the noise reduction achieved through multiple non-destructive samples can now be evaluated.
The curve in Fig. \ref{fig:noiseAtFermi} shows the standard deviation for a series of images with different numbers of samples per pixel, $N_S$. The comparison shows that the mixed image follows the theoretical trend of Eq. \ref{ec:noise_reduction} as expected.

\subsection{Node Removal Inefficiency}
\label{subsec:NRI-FW-SC}
Although conventional CCD devices use a fixed reference voltage to reset the sense node and measure the pedestal level before transferring the charge, using a clock to control the potential applied to the reference voltage ($V_{\text{ref}}$ in Fig. \ref{fig:mas-schematic}) has been shown to improve the NRI \citep{irigoyen2025multiple}. Once the charge readout is complete and the charge is ready to be returned to the extended serial register, $V_{\text {ref}}$ is switched to a higher value, generating a larger potential gradient relative to the PS gate. This change establishes a steeper gradient directed toward the H1 gate, thereby improving the extraction of the charge packet. NRI serves as a metric for measuring the efficiency of this charge removal process from the sense node (SN).

One method to approximate this inefficiency in a low-NRI regime is to evaluate what fraction of the charge in the pixels of the last column of the active area spills into the first column of the overscan region \citep{irigoyen2025multiple}. This phenomenon occurs progressively as the charge packet advances through the amplifier chain. However, when the NRI is small, the contribution from CTI to the measured charge in the first overscan column may become comparable to that of the NRI. The two contributions can be disentangled by exploiting the dependence of the NRI on the amplifier index, whereas the CTI contribution is assumed to be independent of it.

To measure this, the device was exposed to a sequence of illumination levels ranging from low charge up to the full-well capacity of approximately $50\,\mathrm{ke^-}$. At each exposure level, a ratio $q_i/Q$ was computed following the procedure described in
\cite{irigoyen2025multiple}, where $q_i$ is obtained from averaging 720 pixels from the first overscan column of the output image from channel $i$ and $Q$ is calculated by averaging 720 pixels from the last column of the active region from the first amplifier. The zero charge baseline is calculated row-by-row as the mean of ten overscan columns from the same row, starting from the fourth overscan column to avoid residual charge.

In the low-NRI regime we can ignore the recursive nature of the NRI and CTI (e.g., \cite{irigoyen2025multiple}) and instead assume that the measured ratio $q_i/Q$ can be approximated as the sum of these two contributions,

\begin{equation}
    \frac{q_i}{Q}
    = \frac{q_{\mathrm{NRI},i}+q_{\mathrm{CTI}}}{Q}
    = \alpha(i-1)+\beta_{\mathrm{CTI}},
    \label{eq:nri_cti}
\end{equation}
with $1 \leq i \leq N_a$, where $\alpha$ represents the incremental NRI contribution per amplifier and $\beta_{\mathrm{CTI}}$ represents the accumulated CTI contribution from the active area. As discussed in \cite{Lapi2024MAS16}, the first amplifier is not affected by the NRI, and therefore the NRI contribution increases with the number of preceding amplifiers. We note that this estimate of the NRI is itself an approximation that assumes that the CTI only contributes in the serial register of the active area. In reality, CTI will also contribute when the charge is passed through the extended MAS serial register. The effect will be to increase the charge in the first overscan pixels as the packet is read by subsequent amplifiers. A more detailed analysis could include the contribution of the CTI along the extended MAS serial register; however, for the time being, we choose the simple modeling approach. For this reason, the value of the NRI that we quote here should be taken as an upper limit on the true NRI contribution.

The average fractional charge deferred into the first overscan column is then calculated as

\begin{equation}
    \frac{\bar{q}}{Q}
    = \frac{1}{N_a}\sum_{i=1}^{N_a}\frac{q_i}{Q}
    = \beta_{\mathrm{CTI}}+\varepsilon_{\mathrm{avg}},
    \label{eq:e_avg}
\end{equation}

where

\begin{equation}
    \varepsilon_{\mathrm{avg}}
    = \frac{1}{N_a}\sum_{i=1}^{N_a}\alpha(i-1)
    = \alpha\frac{N_a-1}{2},
    \label{eq:nri_avg}
\end{equation}

represents the average NRI contribution across all output amplifiers and quantifies its impact on the final averaged images.
\begin{figure}[b!]
\centering
\begin{subfigure}[t]{0.48\linewidth}
    \centering
    \includegraphics[trim={0 0 0 0}, clip=true, width=\linewidth]{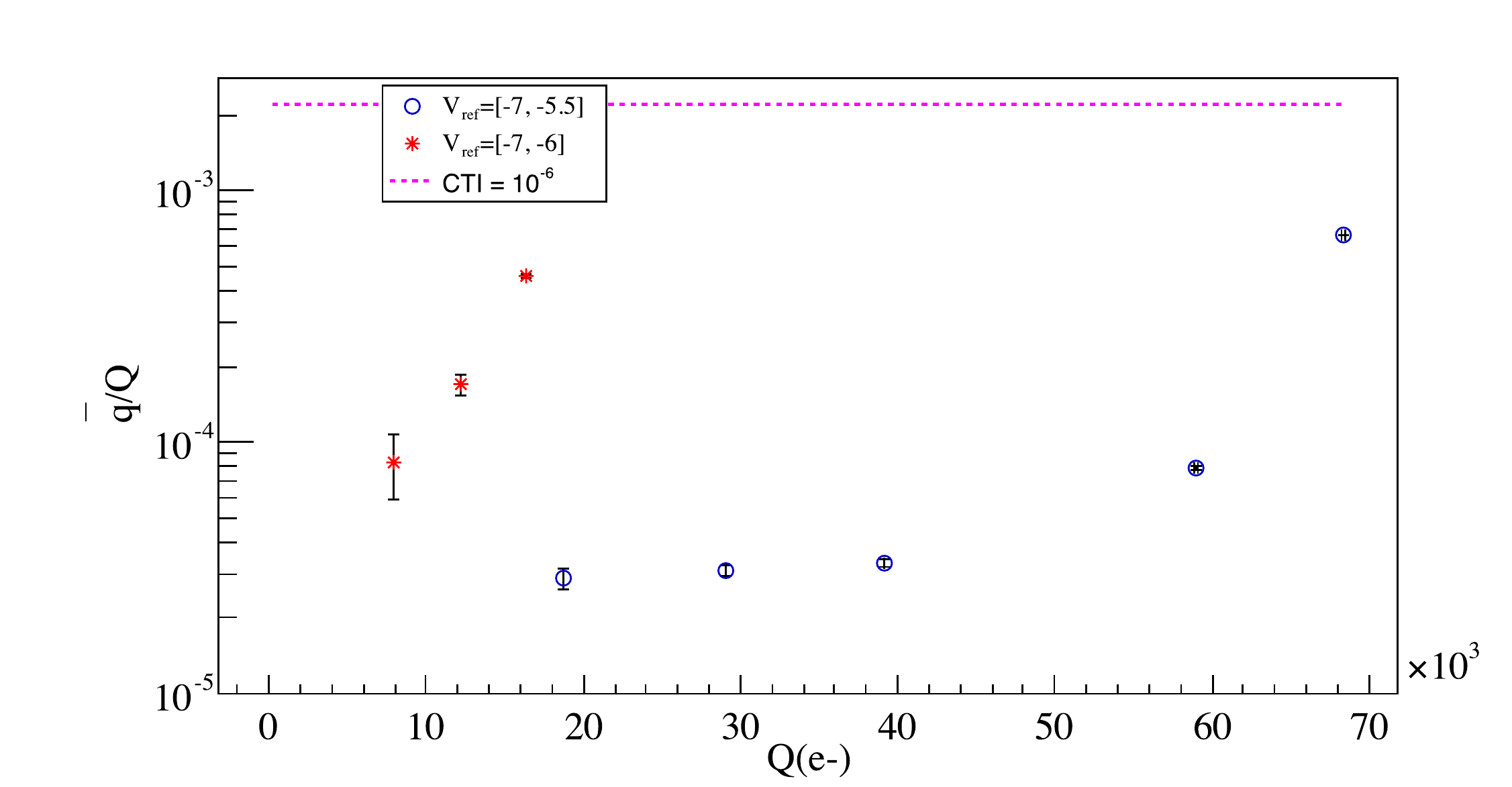}
    \caption{}
    \label{fig:nri_at_fnal_a}
\end{subfigure}
\hfill
\begin{subfigure}[t]{0.48\linewidth}
    \centering
    \includegraphics[trim={0 0 0 0}, clip=true, width=\linewidth]{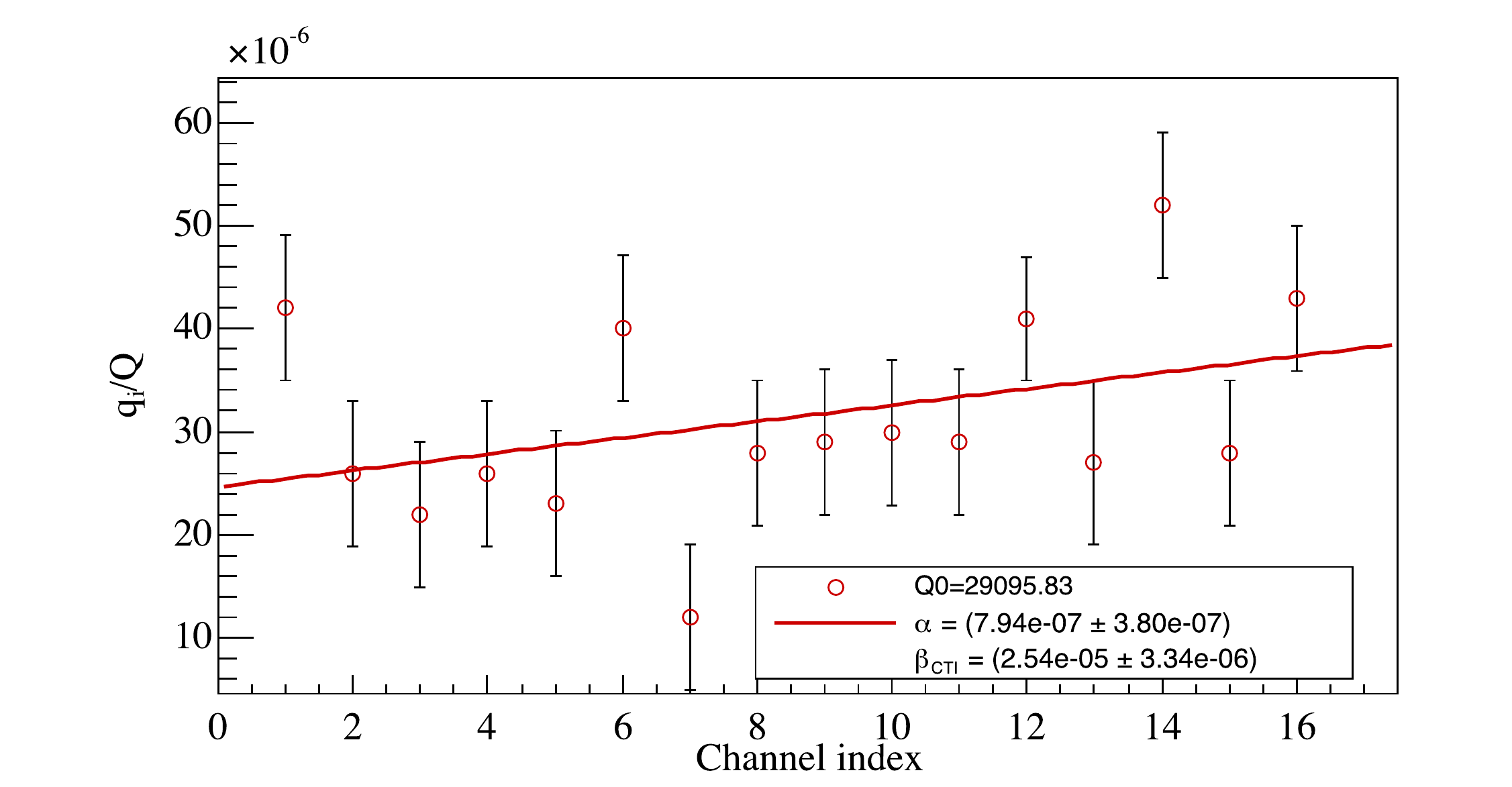}
    \caption{}
    \label{fig:nri_at_fnal_b}
\end{subfigure}
\caption{(a) Average fractional charge deferred into the first overscan column ($\bar{q}/Q$), including both CTI and NRI contributions, as a function of collected charge for the characterized sensor, for 1 volt and 1.5 volt $V_{\text{ref}}$ swings, indicated in red and blue, respectively. The dashed pink line represents the charge distortion equivalent to a CTI of $10^{-6}$ per pixel transfer. (b) Individual $q_i/Q$ terms measured for each channel for a charge packet of approximately $29.1\,\mathrm{ke^-}$, obtained by averaging 720 rows. The solid red line represents a linear fit following Eq.~\ref{eq:nri_cti}, where the slope $\alpha$ represents the incremental NRI contribution per amplifier and $\beta_{\mathrm{CTI}}$ represents the accumulated CTI contribution.}
\label{fig:nri_at_fnal}

\end{figure}

Fig.~\ref{fig:nri_at_fnal_a} shows the resulting $\bar{q}/Q$ values for two different $V_{\text {ref}}$ voltage configurations. The dashed pink line represents a reference level corresponding to the charge distortion produced by a CTI of $10^{-6}$ per pixel transfer for the 512 pixels in the horizontal register. Both curves in the plot show an increasing trend in charge packet size. The impact of the high NRI at high charge levels was one of the main driving factors for voltage optimization. The red dots, with the smaller swing in $V_{\text {ref}}$, reach the maximum allowed distortion for charge packets smaller than the FW requirement. The blue dots in the plot show how the NRI improved as the $V_{\text{ref}}$ swing was increased by half a volt. In this case, the measured charge distortion, including both CTI and NRI contributions, remains below the reference level for the entire charge range of interest.

Figure \ref{fig:nri_at_fnal_b} presents the individual terms $q_i/Q$ from Eq.~\ref{eq:e_avg} as a function of the channel index, $i$, for a charge packet of approximately $29.1\,\mathrm{ke^-}$, which represents a charge value in the mid-FW range. To improve the statistical precision of the measurement, 720 pixels from the corresponding columns were averaged. A linear fit to the data following Eq.~\ref{eq:nri_cti} yields $\alpha=(7.94\pm3.80)\times10^{-7}$ and $\beta_{\mathrm{CTI}}=(2.54\pm0.33)\times10^{-5}$. For the 16-amplifier device, the extracted upper limit on the NRI corresponds to an average contribution of $\varepsilon_{\mathrm{avg}}=(5.96\pm2.85)\times10^{-6}$ to the final averaged image. Normalizing the CTI contribution by the number of horizontal transfers (N=512) yields a CTI of $(4.96\pm0.65)\times10^{-8}$ per pixel transfer. The small slope obtained from the fit indicates that the measured charge distortion at this charge level is predominantly associated with CTI rather than NRI. In fact, the charge increase measured from amplifier 1 to amplifier 16 is completely consistent with what would be expected from CTI alone in the extended serial register.

\subsection{Full-Well capacity}
The FW capacity was studied using the standard Photon-Transfer Curve (PTC) method \citep{janesick2001scientific, janesick2007photon}. In theory, the PTC follows a slope of $1/2$ on a log--log scale due to shot noise. The FW is defined as the point where this trend abruptly breaks, indicating that the potential well is saturated and charge begins to spread into adjacent pixels.

\begin{figure}[h!]
    \centering
    \includegraphics[width=.8\linewidth]{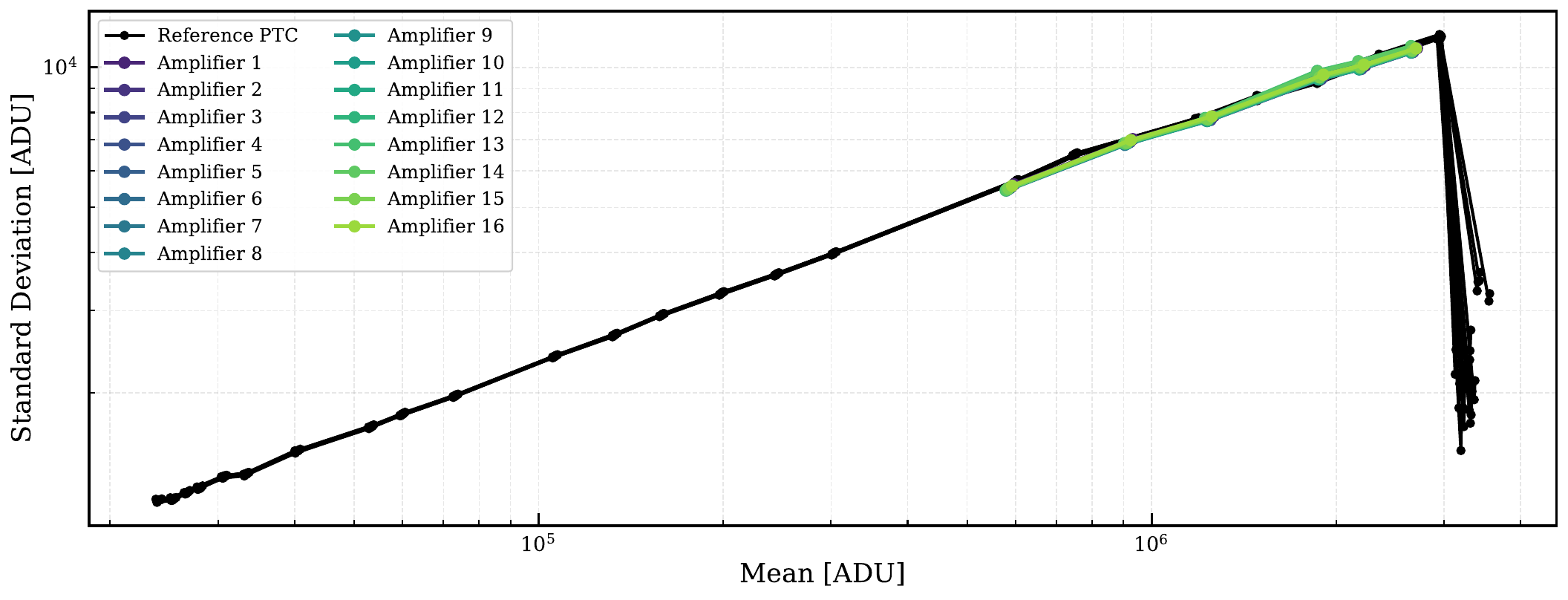}
    \caption{Photon-Transfer Curve measurement on MAS-CCD sensors. The black line corresponds to measurements from a reference MAS-CCD that was extensively characterized, while the green lines corresponds to the commissioned detector. These curves show good agreement with the reference, which has a FW of 57186 e$^{-}$.}
    \label{fig:PTC@FNAL}
\end{figure}

Figure~\ref{fig:PTC@FNAL} shows the PTC measured for each amplifier for two sensors. The black curve corresponds to a full-range PTC obtained from the most extensively studied device, which served as the reference during the FW--SC optimization process. The colored curves show the PTC per amplifier for the commissioned device, measured over the signal range of interest. As expected, these curves overlap with the reference curve. Since the PTC turnover was not measured directly for the commissioned device, the same threshold is assumed for both sensors.
The highest mean value before the break point is extracted from the PTC of the reference device and converted using the system conversion gain ($52.46$ ADU/e$^-$), yielding an approximate FW of $\sim57\mathrm{ke}^-$.

\subsection{Spurious Charge}

The optimization task results in a fine-tuning of the trade-off between FW and horizontal SC. SC measurements are usually challenging, as low-noise data are required to disentangle its mean value from the electronic noise. Therefore, the horizontal SC was measured using $N_S = 36$ after averaging the output of all amplifiers. The measured value after the optimization was \meg{$1.01 \times 10^{-3} \, $e$^-/\text{pix}/\text{transfer}$}, meeting the requirement established to ensure that the spurious charge (SC) contribution to the pixel uncertainty remains comparable to that of the readout noise after averaging the signals from the 16 amplifiers in a single-sample readout. The optimization approach was based on the horizontal spurious charge as the metric, given that the horizontal clocks exhibit the largest swings to handle large charge packets at a much faster speed than the vertical clocks.

\subsection{Summary and other parameters}
The optimized voltages are presented in Table \ref{tab:voltages}, under the column ``High pixel signal voltages". Additionally, the table presents a set of low-noise voltages, typically used in laboratory measurements, under the column ``Low pixel signal voltages". This configuration inherited from previous optimizations delivers a lower FW and a smaller spurious charge contribution, and it was used for a single test where the pixels have very short exposure to light, as detailed in the next section.

\begin{table}[h!]
\centering
\begin{tabular}{lllll}
\hline
\multicolumn{1}{c}{Signal name}                         & \multicolumn{2}{c}{High pixel signal voltages}  & \multicolumn{2}{l}{Low pixel signal voltages} \\ \hline
\multicolumn{1}{l|}{}                                   & Low {[}V{]} & \multicolumn{1}{l|}{High {[}V{]}} & Low {[}V{]}           & High {[}V{]}          \\
\multicolumn{1}{l|}{Vertical (V)}                       & -3.0        & \multicolumn{1}{l|}{4.0}          & -3.5                  & 1.5                   \\
\multicolumn{1}{l|}{Horizontal (H)}                     & -9.9        & \multicolumn{1}{l|}{2.7}          & -9.5                  & -2.0                  \\
\multicolumn{1}{l|}{Reference Voltage $V_{\text{ref}}$} & -7.0        & \multicolumn{1}{l|}{-5.5}         & -7.0                  & -5.5                  \\
\multicolumn{1}{l|}{Summing Well (SW)}                  & -9.9        & \multicolumn{1}{l|}{2.5}          & -9.9                  & -3.0                  \\
\multicolumn{1}{l|}{Output Gate (OG)}                   & -8.0        & \multicolumn{1}{l|}{0.0}          & -8.0                  & -4.0                  \\
\multicolumn{1}{l|}{Dumping Gate (DG)}                  & -10.0       & \multicolumn{1}{l|}{2.0}          & -10.0                 & -2.0                  \\
\multicolumn{1}{l|}{Pixel Separation (PS)}              & -8.9        & \multicolumn{1}{l|}{1.9}          & -8.0                  & -2.0                  \\ \hline
\end{tabular}

   \caption{Two sets of voltages for high and low pixel signal performance, used for MAS-CCD on-sky operation.}
   \label{tab:voltages}
\end{table}

A summary of the requirements and main CCD parameters are presented in Table \ref{table:parameters}. Other requirements also listed are the dark-current and maximum charge diffusion based on constraints from similar fully depleted CCDs used for the Dark Energy Camera \citep{flaugher2015dark}. In addition to these parameters, the table reports the pixel time for single-sample operation and the additional time required for each extra sample taken.

\begin{table}[h!]
\centering
\begin{tabular}{lllll}
\hline
CCD parameters & Requirement/Value   & \multicolumn{2}{c}{ Measured value}  & Units      \\ \hline
  &       & High pixel signal     & Low pixel signal &  \\ \hline
Full Well                        & \textgreater{}50000                    & 57186   & 10000    & e$^-$      \\
Horizontal spurious charge  & \textless{}$1.95\times10^{-3}$     & \textless{} $1.01\times10^{-3}$ & \textless{} $5.8\times10^{-4}$       & e$^-$/pix/transfer \\
Dark Current                     & \textless{}25 & 17.48    & 7.18      & e$^-/$pix$/$h    \\
CTI                              & \textless{}$1\times10^{-6}$   & \textless{}$4.96\times10^{-8}$   &  \textless{}$1.32\times10^{-6}$    & $/$transfer \\
Average NRI                      & \textless{}$2.22\times10^{-3}$         & $5.96\times10^{-6}$         & \textless{}$6.79\times10^{-4}$   & e$^-/$transfer   \\
Readout noise amplifier averaging for $N_S=1$ & \textless{}1.5  & 1.16     & 1.11     & e$^-$      \\
Lateral diffusion                & \textless{}7.5     & \textless{}7.5  & \textless{}7.5   & $\mu m$    \\
\hline
Pixel time with $N_S=1$          & 15.5  & - & - & $\mu$s  \\
Extra sample time                & 13.46  & -   &  - & $\mu$s  \\
Gain                             & 52.46    & -  &  -   & ADU/e$^-$    \\
CCD Form factor                        & 1024$\times$512     & -    &  -    &   pixels   \\
Pre-scan pixels                         & 27     & -   & -   & pixels     \\
Inter-amplifiers pixels            & 15     & -  & -  &  pixels   \\
Pixel size                       & $15\times15$     & -  & -  &  $\mu$m   \\
Thickness                       & $250$     & -  & -  &  $\mu$m   \\
Operating temperature                       & $140$ to $160$    & -  & -  &  K   \\
\hline
\end{tabular}
\caption{CCD parameters and performance requirements used for the voltage optimization, together with the measured sensor characteristics for two operating configurations. The low pixel signal configuration was used during characterization, but does not satisfy the full-well requirement, whereas the optimized high-pixel-signal configuration meets the required performance limits.}
\label{table:parameters}
\end{table}

\section{Camera integration}
\label{sec:camera_integration}

The sensor was commissioned on the SMARTS–GSU 0.9-meter telescope, an optical reflecting Cassegrain telescope located at the Cerro Tololo Inter-American Observatory (CTIO) in Chile, at an altitude of 2241 m. The telescope uses a Cassegrain optical design and is mounted on an equatorial mount inside a hemispherical dome. A CCD-based autoguider, which looks at a small off-axis field, is permanently installed on the telescope. Standard \textit{BVRI} and \textit{ugriz} filter sets were made available by NSF AURA.
The active area of the sensor (1024 $\times$ 512 array of 15-$\mu$m pixels) provides a Field of View (FOV) of approximately $4.34 \times 2.17$ arcmin$^2$, with a pixel scale of $0.2546$ arcsec$/$pixel.

\begin{figure}[t]
\centering
\begin{subfigure}[t]{0.57\linewidth}
    \centering
    \includegraphics[trim={0 0 0 0}, clip=true, width=\linewidth]{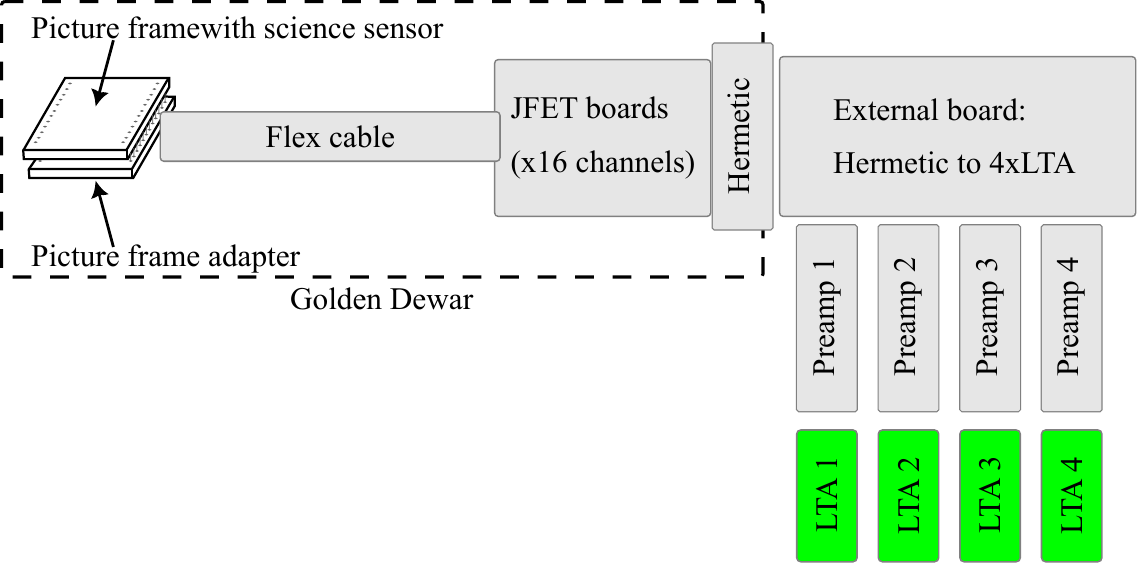}
    \caption{}
    \label{fig:archi_a}
\end{subfigure}
\hfill
\begin{subfigure}[t]{0.40\linewidth}
    \centering
    \includegraphics[trim={0 0 0 0}, clip=true, width=\linewidth]{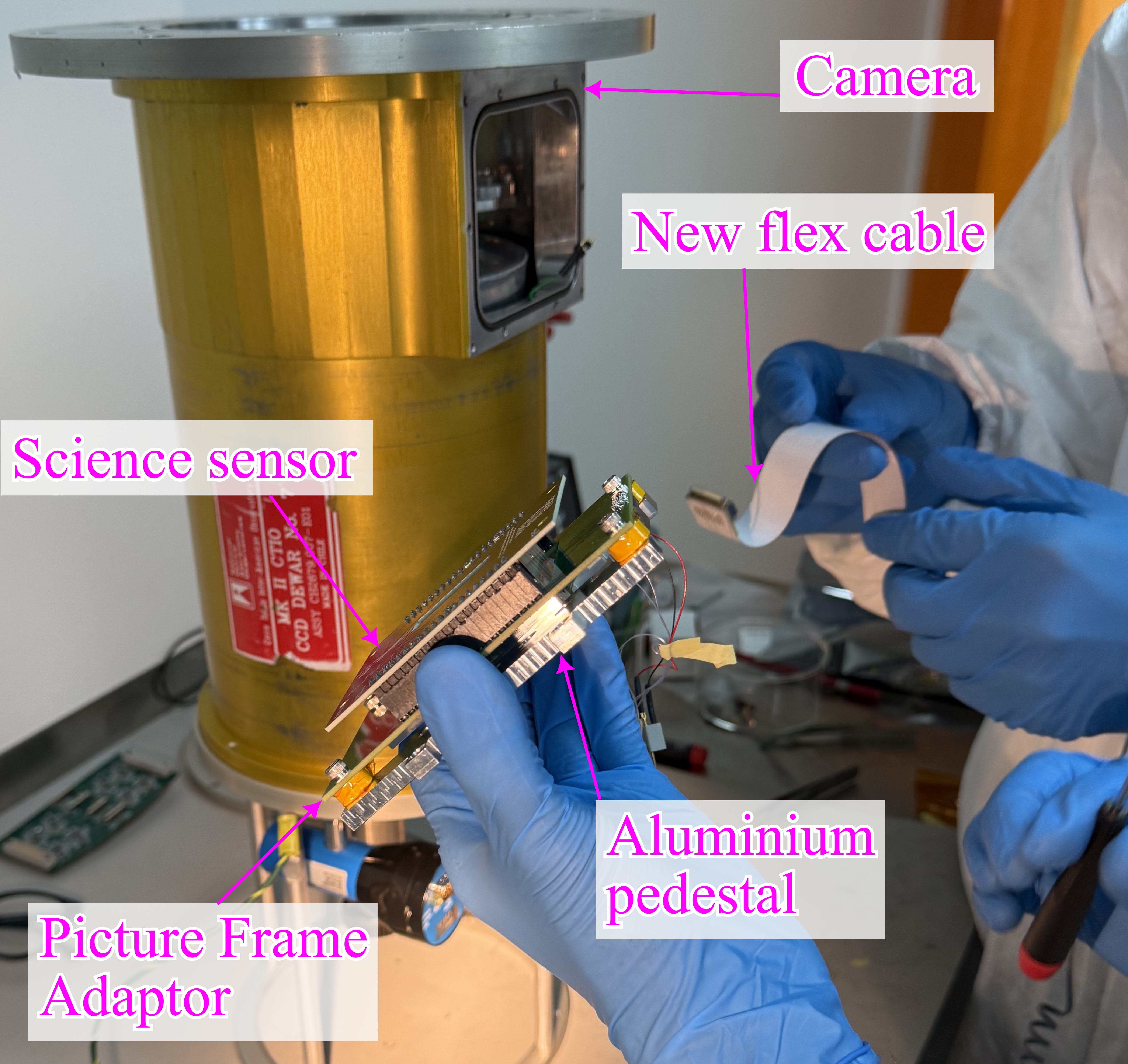}
    \caption{}
    \label{fig:archi_b}
\end{subfigure}
\caption{(a) Final camera architecture. (b) Photograph of the MAS-CCD camera integration at the CTIO 0.9-m telescope. The principal parts, including the science sensor, the aluminum pedestal, the flex cable, camera, and picture frame adaptor, are labeled.}
\label{fig:archi}

\end{figure}

The NOIRLab/AURA team led the refurbishment of a Dewar and the mechanical and electrical integration of the MAS-CCD into it. Figure~\ref{fig:archi_a} shows the final camera architecture. This work included incorporating the new mechanical mount for the sensor and designing, fabricating, and installing the compact interface electronics and interconnects required for the camera. These components included the flex cable, internal JFET board, hermetic interface, and external motherboard connecting the preamplifiers to the LTA controllers. The camera electronics integration was tested to be fully functional with an engineering-grade sensor. Figure~\ref{fig:archi_b} shows a picture during the installation of the science-grade sensor. The team also integrated the LTA readout software with the telescope data-acquisition and control system, enabling on-sky operation of the sensor. This system was the same as that used to test Skipper-CCDs at SOAR \citep{bonati2024skipper}.

\begin{figure}[t]
\centering
\begin{subfigure}[t]{0.42\linewidth}
    \centering
    \includegraphics[trim={0 0 0 0}, clip=true, width=\linewidth]{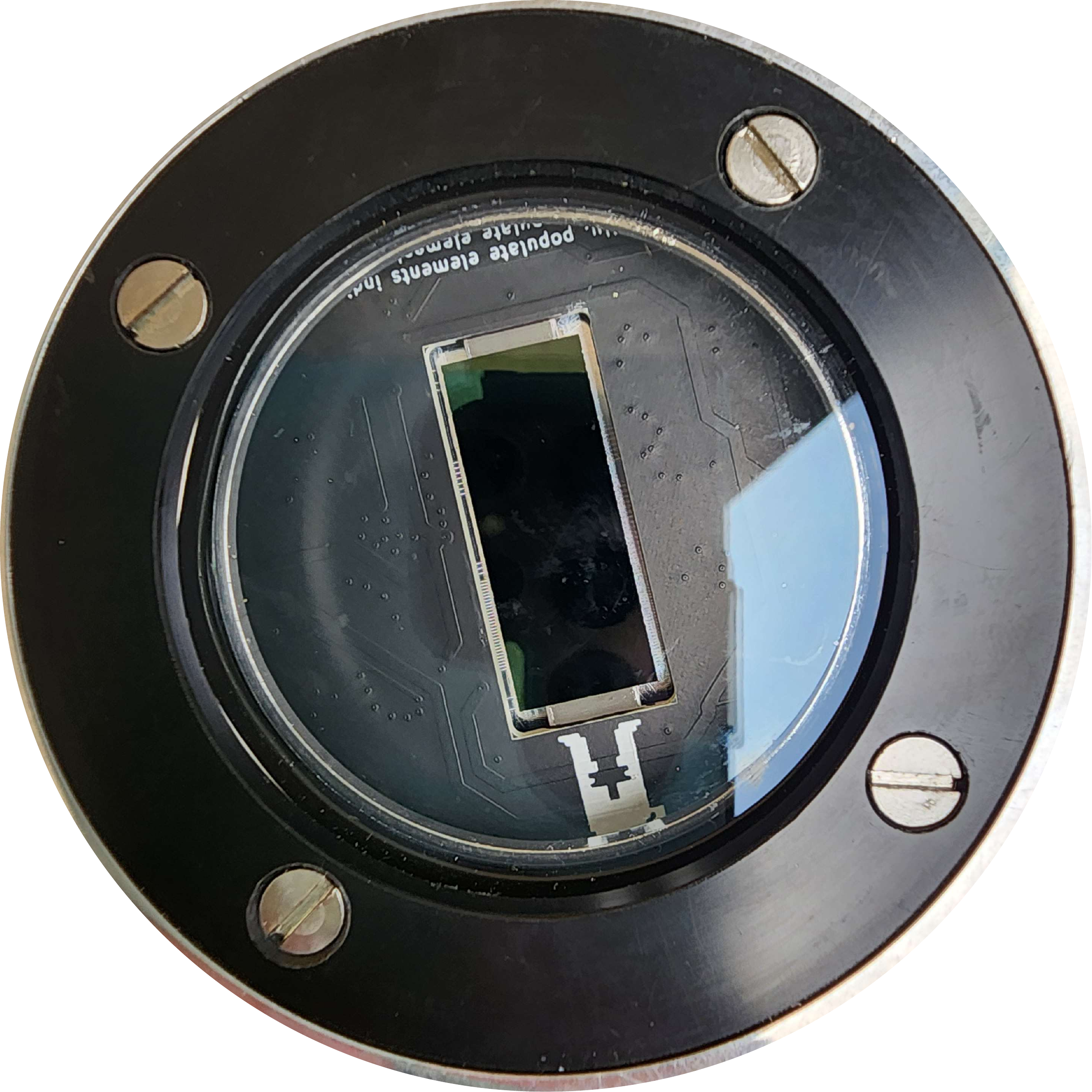}
    \caption{}
    \label{fig:telescope_a}
\end{subfigure}
\hfill
\begin{subfigure}[t]{0.48\linewidth}
    \centering
    \includegraphics[trim={0 0 0 0}, clip=true, width=\linewidth]{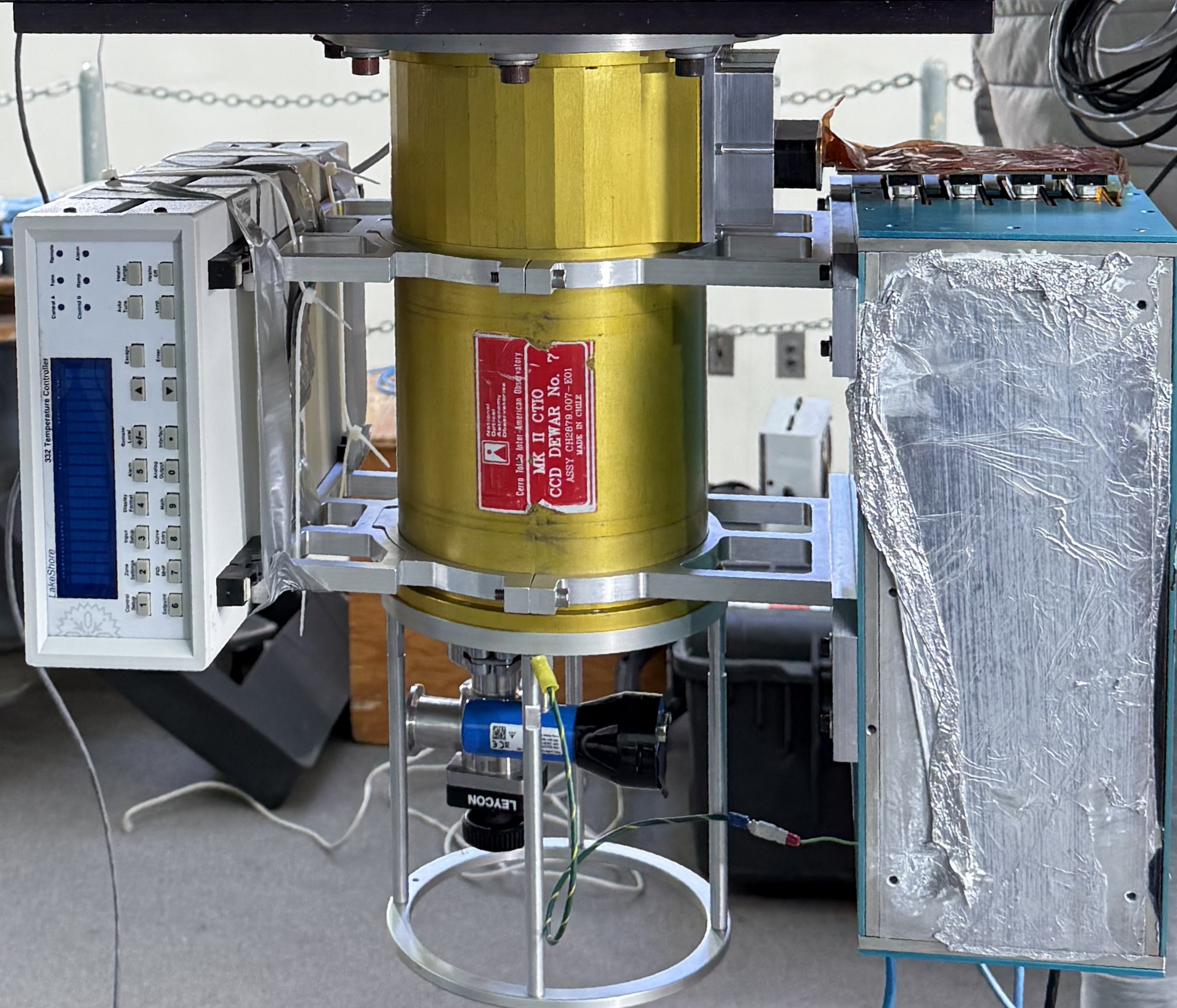}
    \caption{}
    \label{fig:telescope_b}
\end{subfigure}

\begin{subfigure}[t]{0.35\linewidth}
    \centering
    \includegraphics[trim={0 0 0 0}, clip=true, width=\linewidth]{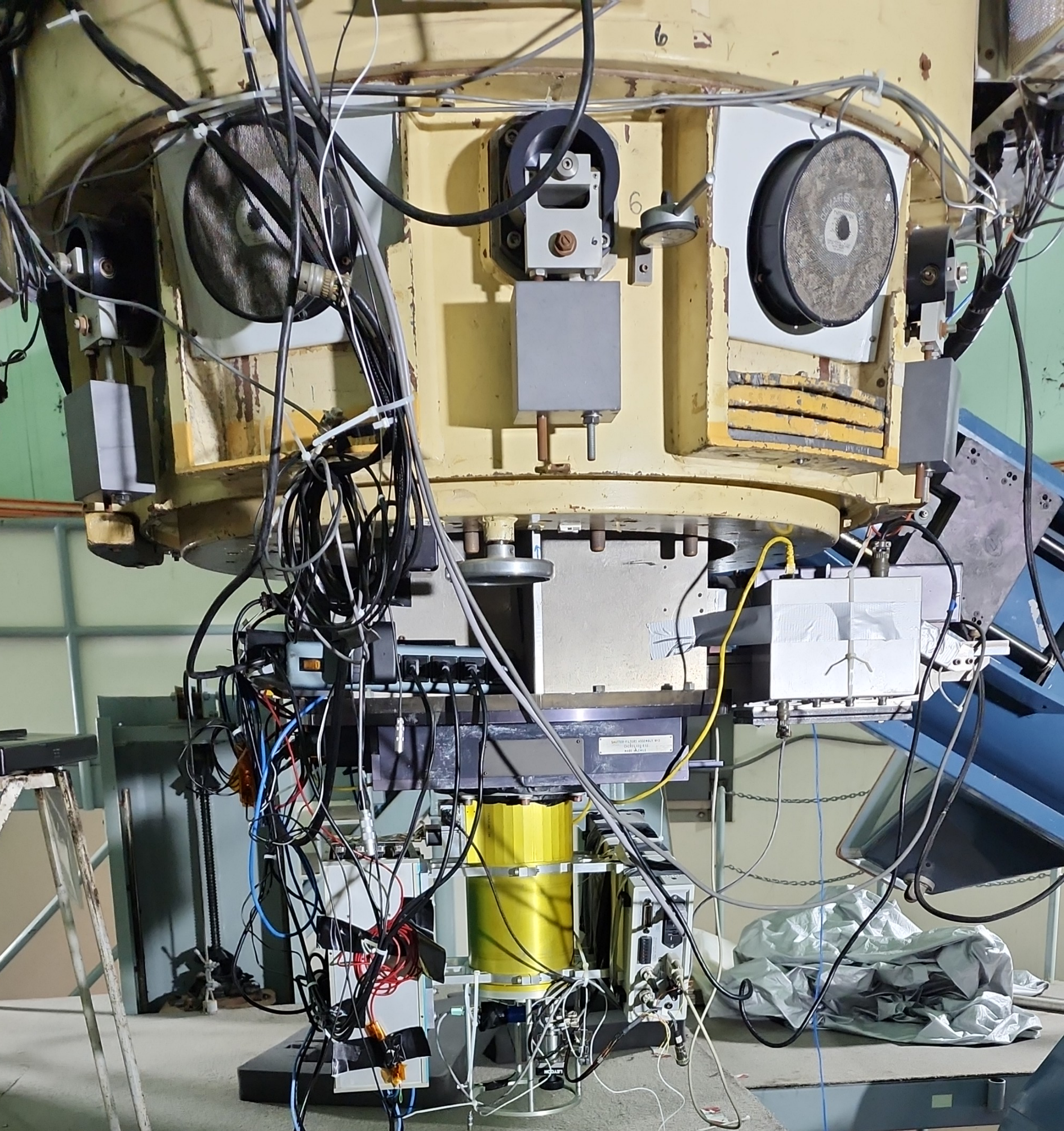}
    \caption{}
    \label{fig:telescope_c}
\end{subfigure}
\hfill
\begin{subfigure}[t]{0.55\linewidth}
    \centering
    \includegraphics[trim={0 0 0 0}, clip=true, width=\linewidth]{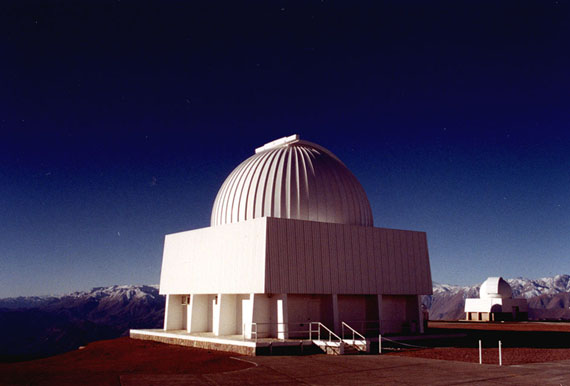}
    \caption{}
    \label{fig:telescope_d}
\end{subfigure}
\caption{(a) Front-facing view of the back-illuminated MAS-CCD through the Dewar window. (b) Close-up side view of the Dewar, temperature controller, and attached LTA electronics enclosure. (c) Integrated camera system mounted at the Cassegrain focus of the telescope. (d) SMARTS–GSU 0.9-meter telescope dome as seen from the exterior.}
\label{fig:telescope}

\end{figure}

Figure~\ref{fig:telescope} presents four photographs illustrating different stages and views of the resulting camera system: \ref{fig:telescope_a} shows a front-facing view of the back-illuminated MAS-CCD installed inside the camera, as seen through the dewar window; the photograph~\ref{fig:telescope_b} provides a close-up side view of the camera, including the Dewar, the temperature controller, and the LTA electronics enclosure attached to the right side of the camera; the photograph \ref{fig:telescope_c} shows the fully integrated camera system mounted on the CTIO 0.9-m telescope; and finally, the dome where the SMARTS–GSU 0.9-meter telescope is located is depicted in Figure~\ref{fig:telescope_d}.

\section{MAS-CCD on sky characterization}
\label{sec:engineering_on_field_characterization}
The on-sky observations were collected between 2025 July 14 and 2025 July 28. For the observation runs presented in the following subsections, the high pixel signal voltage configuration was used, except for the results presented in subsection \ref{sec:continuous_readout}, where low pixel signal voltages were used.

The following subsections are divided as follows: first, the noise performance in the telescope environment is described and characterized. Then, the impact of the NRI on the observations, along with the validation of the NRI mitigation technique, is presented. Section \ref{sec:roi} shows on-sky performance results for two readout modes that allow increasing the readout speed by ignoring unnecessary pixels in the readout scheme. This is followed by a case study of system robustness and stability through high-cadence observation of a variable star. Finally, the section concludes by presenting multi-channel false color images of two nebulae observed with the MAS-CCD.

\subsection{Noise on sky}
\begin{figure}[t]
\centering
\begin{subfigure}[t]{0.32\linewidth}
    \centering
    \includegraphics[trim={0 0 0 0}, clip=true, width=\linewidth]{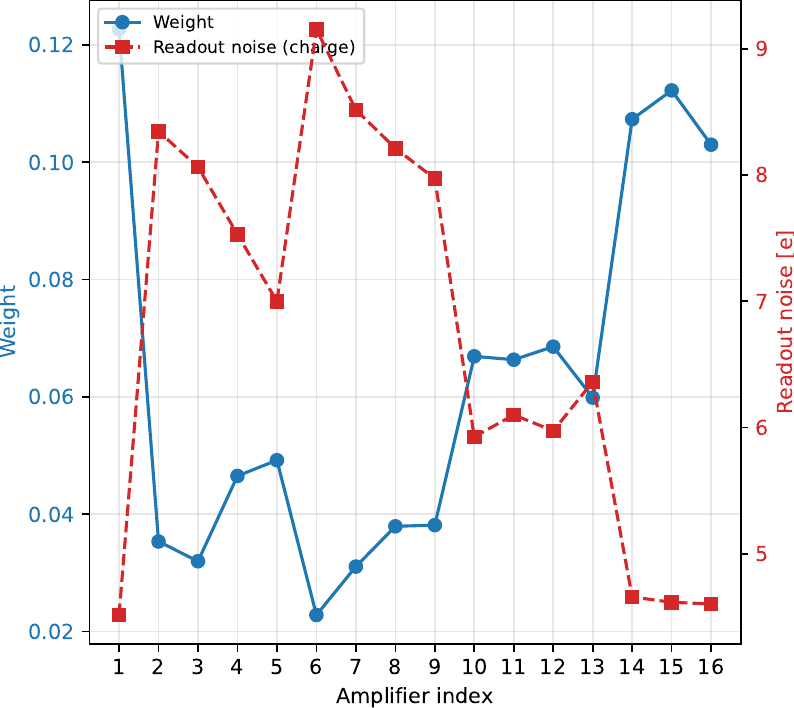}
    \caption{}
    \label{fig:atCTIO_a}
\end{subfigure}
\hfill
\begin{subfigure}[t]{0.32\linewidth}
    \centering
    \includegraphics[trim={0 0 0 0}, clip=true, width=\linewidth]{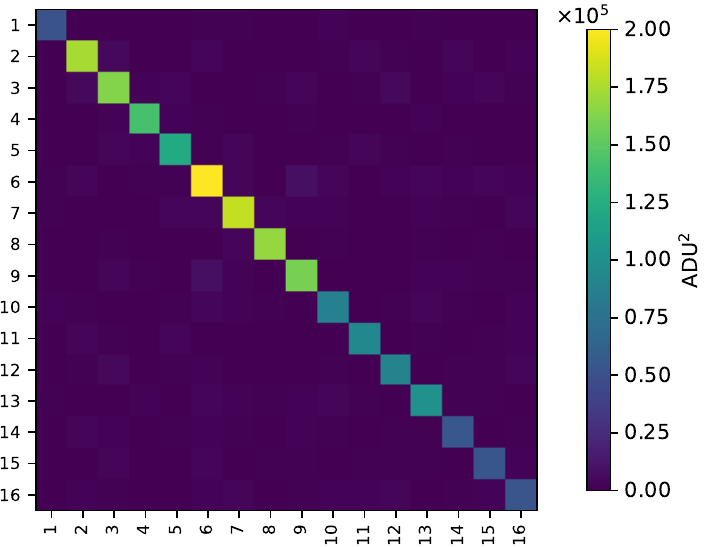}
    \caption{}
    \label{fig:atCTIO_b}
\end{subfigure}
\hfill
\begin{subfigure}[t]{0.32\linewidth}
    \centering
    \includegraphics[trim={0 0 0 0}, clip=true, width=\linewidth]{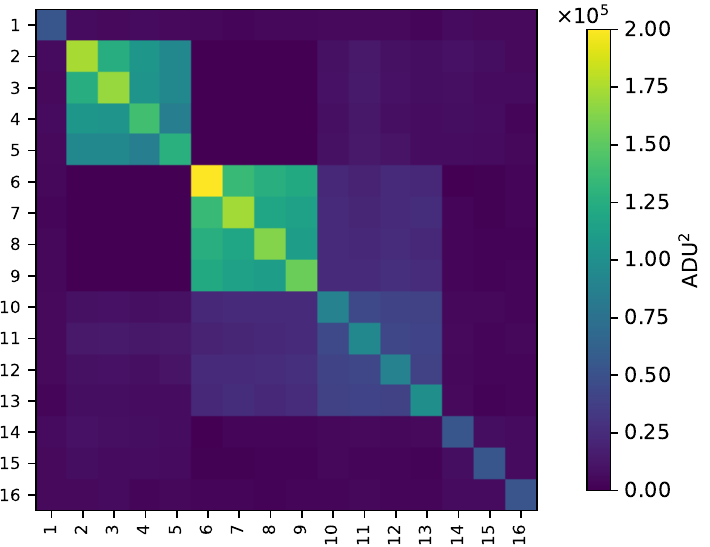}
    \caption{}
    \label{fig:atCTIO_c}
\end{subfigure}

\caption{Summary of performance for one sample per pixel ($N_S = 1$) at CTIO: (a) Standard deviation of the readout noise (red squares) and the optimal weights (blue dots) to average the information from the different channels; (b) correlation matrix of the readout noise among different channels for pixels in charge phase, and (c) similar matrix for the pixels in time phase. }
\label{fig:atCTIO}

\end{figure}

Changes in grounding conditions and readout electronics between the Fermilab laboratory and CTIO telescope configurations can alter the noise properties of the individual channels and their correlations. Moreover, four different LTA controllers were used. Therefore, the noise characterization was repeated at CTIO, and the optimal channel weights were recalculated for the on-sky configuration.
A summary of the noise performance of the device at the CTIO telescope is presented in Fig. \ref{fig:atCTIO}. In this case, the individual noise ranges between $4.5\,$e$^-$ and $9.0\,$e$^-$, showing a larger noise contribution from some of the channels compared to previous characterization in Section \ref{sec:noiseatFermi}. Therefore, the resulting optimal weights differ. Table \ref{tab:routing_ctio} shows the physical channel routing to the LTA controllers. Channels 1, 14, 15, and 16 achieve lower noise than those observed in Section \ref{sec:noiseatFermi} characterized at Fermi and correspond to the leader LTA. This suggests some noise issue affecting the controllers individually. This is supported by the time-phase correlated matrix in Fig.~\ref {fig:atCTIO_c} that shows higher diagonal values for those channels and high correlation with the channels belonging to the same follower LTA. Since the charge-phase covariance matrix in Fig.~\ref {fig:atCTIO_b} shows no evident cross-channel correlation, the optimal weights do little to suppress the extra noise besides weighting the noisier channels less.

\begin{table}[h!]
    \centering
    \begin{tabular}{|c|c|c|c|c|}
        \hline
        \textbf{LTA number} & 1 & 2 & 3 & 4 \\
        \textbf{Configuration} &  Leader & Follower & Follower & Follower \\
        \textbf{Amplifier index} &1, 14, 15, 16 & 10, 11, 12, 13 & 2, 3, 4, 5 & 6, 7, 8, 9 \\
        \hline
    \end{tabular}
    \caption{Routing of MAS-CCD amplifier outputs to LTA channels at CTIO.}
    \label{tab:routing_ctio}
\end{table}

The readout standard deviation achieved with the leader LTA is a strong indicator that there is still room for improvement in the noise performance of the follower LTA controller boards. Such improvement could likely be achieved through further hardware and grounding optimization. However, the limited observing window favored a conservative risk-management approach, no hardware modifications were made to the system in order to preserve the readiness of the system for data taking during nighttime observations. If needed, further noise reduction remained available through the use of multiple samples per amplifier. Fig. \ref{fig:noiseAtFermi} shows the noise after averaging the channels as a function of $N_S$ in red dots. Even though the single-sample noise is larger than the laboratory measurements, indicated in magenta dots, the expected $\propto 1/\sqrt{N_S}$ reduction is still conserved.

\begin{figure}[h!]
\centering
\begin{subfigure}[t]{0.75\linewidth}
    \centering
    \includegraphics[trim={0 0 0 0}, clip=true, width=\linewidth]{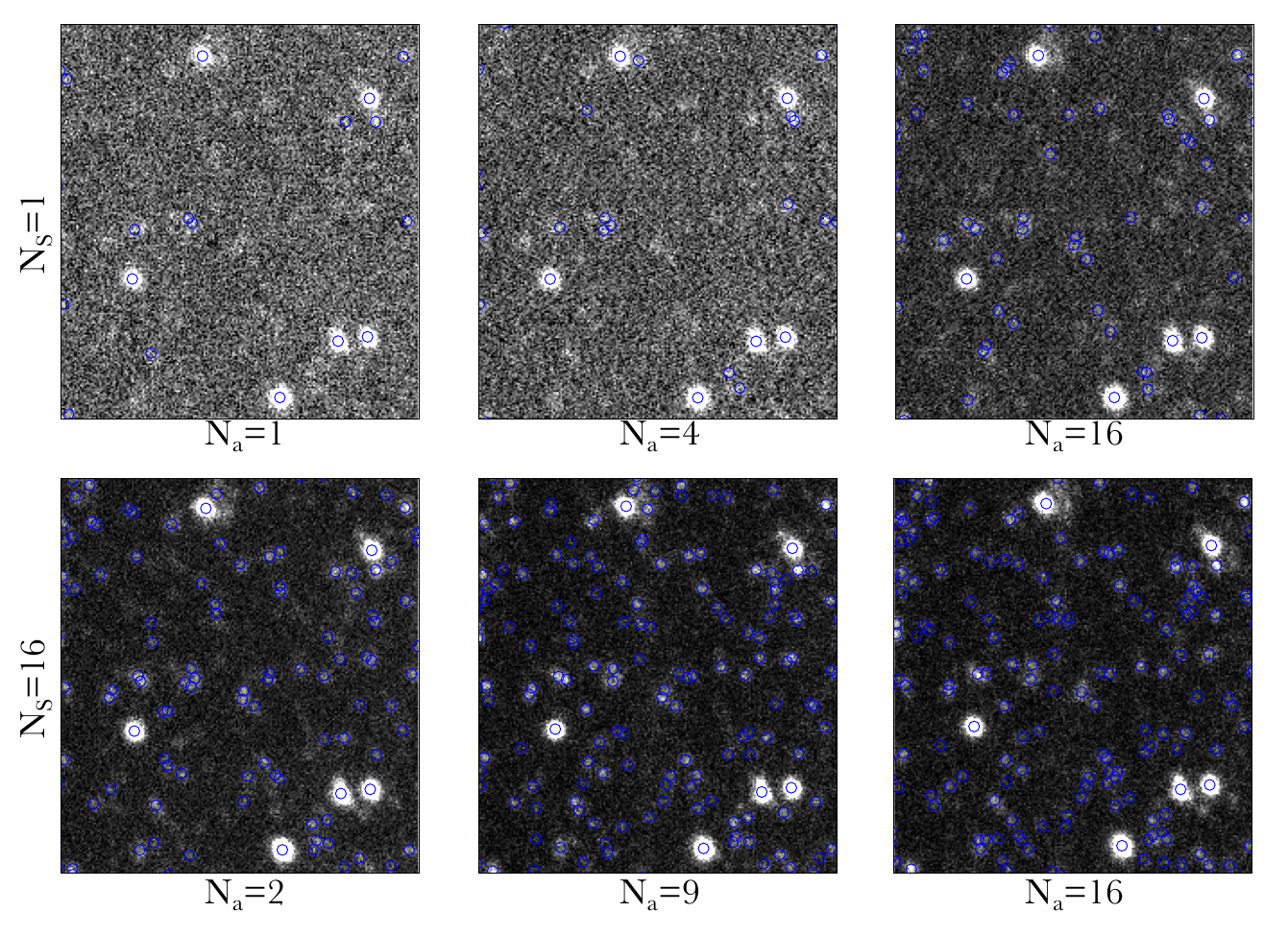}
    \caption{}
    \label{fig:counting_stars_a}
\end{subfigure}
\hfill

\begin{subfigure}[t]{0.45\linewidth}
    \centering
    \includegraphics[trim={0 0 0 0}, clip=true, width=\linewidth]{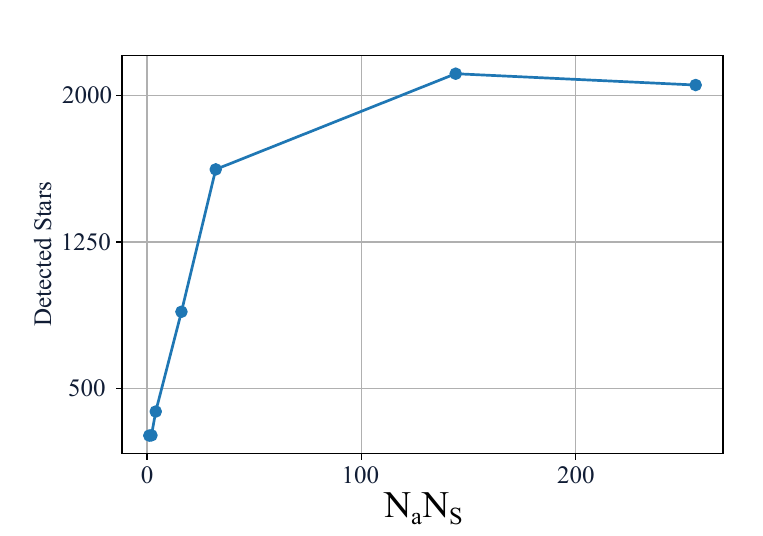}
    \caption{}
    \label{fig:counting_stars_b}
\end{subfigure}
\hfill
\begin{subfigure}[t]{0.45\linewidth}
    \centering
    \includegraphics[trim={0 0 0 0}, clip=true, width=\linewidth]{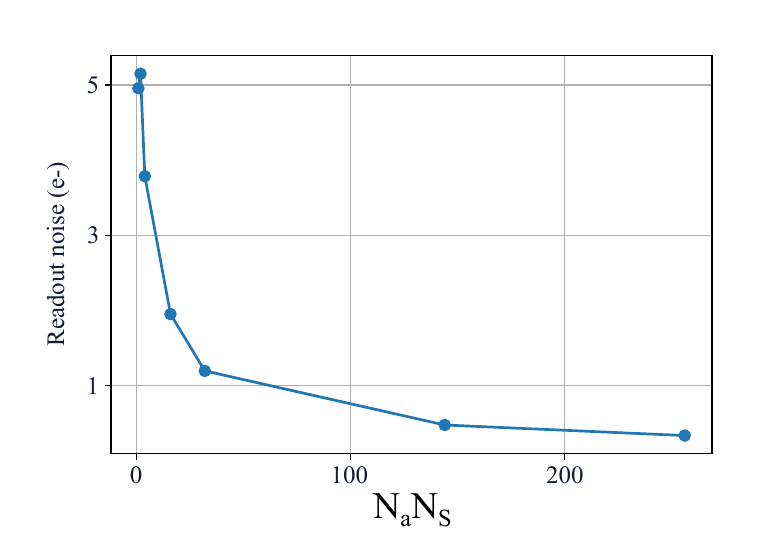}
    \caption{}
    \label{fig:counting_stars_c}
\end{subfigure}
\caption{(a) Cutouts from images taken with different combinations of $N_S$ and $N_a$. The blue circles indicate where the stars were detected by the software. (b) Number of detected stars as a function of the product $N_SN_a$. (c) Readout noise as a function of the product $N_SN_a$. The saturation in the number of detected stars at large $N_SN_a$ suggests that further reductions in readout noise provide diminishing improvements in source detection, as other noise contributions, such as the sky background, may become comparable.
}
    \label{fig:counting_stars}

\end{figure}

To connect this noise reduction with an on-sky observable, a complementary test was performed on a region of the 47 Tucanae globular cluster. A set of images of the same FOV was taken with a 500 ms exposure time using the Johnson-Cousins $V$-band filter, varying the number of amplifiers ($N_a$) and samples per amplifier ($N_S$). Then, these images were processed using SExtractor \citep{Source_Extractor} to detect the number of stars with SNR $\gtrsim 4.5$ (i.e., DETECT MINAREA=5 and DETECT THRESH=2).
Figure \ref{fig:counting_stars_a} shows a small segment of 225 by 200 pixels from the taken images to visualize the impact of the noise reduction. Increasing $N_S$ and $N_a$ increases the number of detected stars by a large amount due to the readout noise reduction of the background pixels.

Figure \ref{fig:counting_stars_b} shows the number of detected stars as a function of the product $N_SN_a$ for the entire field of view. Initially, around 250 stars were detected, and this number increased to over 2000 when $N_SN_a$ product exceeds 100. The corresponding readout noise reduction, measured from the overscan of the images, is detailed in Figure \ref{fig:counting_stars_c}. These plots reveal that a decrease in readout noise from 5 to 1 electron increases the detected stars from 250 to around 1600. This shows that using multiple amplifiers reduces readout noise and improves detection of faint sources that would otherwise remain hidden in the readout noise. The last two points in the plot show that readout-noise reduction saturates the number of detected stars, suggesting that other components of the data, such as the sky background, might contribute comparably in this regime.

\subsection{Effect of the NRI on sky}
As discussed in Section \ref{subsec:NRI-FW-SC}, NRI is an important performance parameter of the MAS-CCD. To evaluate its effect on-sky, and to test whether the proposed mitigation technique remains effective at the telescope, a target star was imaged with a fixed exposure time under two voltage configurations. In the first configuration, $V_{\text {ref}}$ was held fixed at $-7$ V, as in standard CCD operation. In the second configuration, $V_{\text {ref}}$ was clocked, from -7 V to -5.5 V, according to the voltage listed in Table~\ref{tab:voltages}, at the time of charge removal from the sense node.

\begin{figure}[h!]
\centering
\begin{subfigure}[t]{0.33\linewidth}
    \centering
    \includegraphics[trim={0 0 0 0}, clip=true, width=\linewidth]{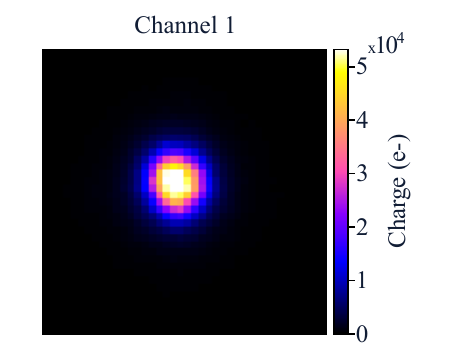}
    \caption{}
    \label{fig:nri_on_sky_a}
\end{subfigure}
\hfill
\begin{subfigure}[t]{0.33\linewidth}
    \centering
    \includegraphics[trim={0 0 0 0}, clip=true, width=\linewidth]{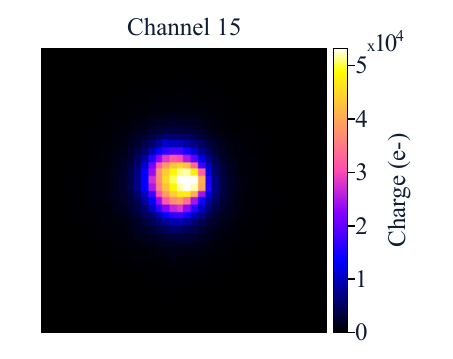}
    \caption{}
    \label{fig:nri_on_sky_b}
\end{subfigure}\hfill
\begin{subfigure}[t]{0.33\linewidth}
    \centering
    \includegraphics[trim={0 0 0 0}, clip=true, width=\linewidth]{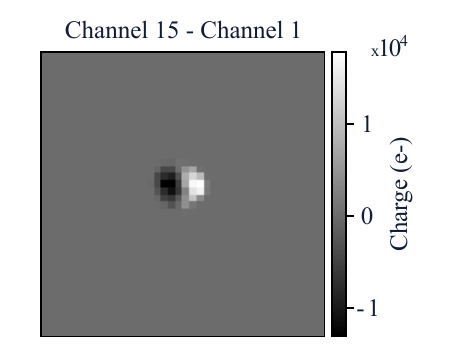}
    \caption{}
    \label{fig:nri_on_sky_c}
\end{subfigure}

\begin{subfigure}[t]{0.33\linewidth}
    \centering
    \includegraphics[trim={0 0 0 0}, clip=true, width=\linewidth]{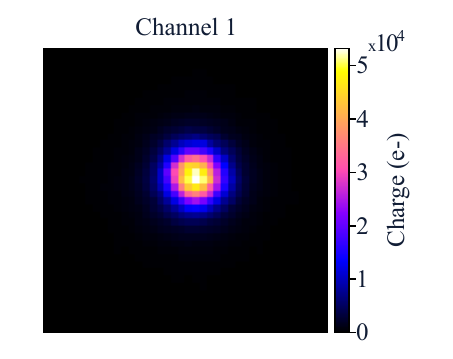}
    \caption{}
    \label{fig:nri_on_sky_d}
\end{subfigure}
\hfill
\begin{subfigure}[t]{0.33\linewidth}
    \centering
    \includegraphics[trim={0 0 0 0}, clip=true, width=\linewidth]{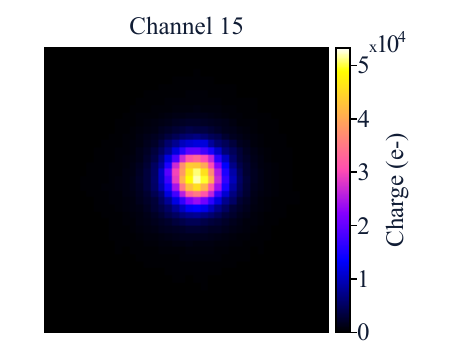}
    \caption{}
    \label{fig:nri_on_sky_e}
\end{subfigure}\hfill
\begin{subfigure}[t]{0.33\linewidth}
    \centering
    \includegraphics[trim={0 0 0 0}, clip=true, width=\linewidth]{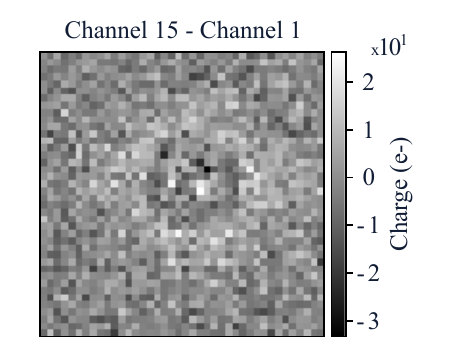}
    \caption{}
    \label{fig:nri_on_sky_f}
\end{subfigure}

\caption{On-sky evaluation of the NRI effect using an 8 s exposure of a star in the NGC 6025 cluster through the Johnson $V$ filter. The top row (a–c) corresponds to the fixed $V_{\mathrm{ref}}=-7$ V configuration, while the bottom row (d–f) corresponds to the clocked $V_{\mathrm{ref}}$ configuration, in which the voltage is switched from $-7$ V to $-5.5$ V during charge removal from the sense node. Panels (a) and (d) show the charge measured by the first amplifier, while panels (b) and (e) show the corresponding measurements by the 15th amplifier. Panels (c) and (f) show the pixel-by-pixel difference between the 15th and first amplifier measurements. The fixed-voltage configuration exhibits a dipole-like residual consistent with charge redistribution caused by NRI, while the clocked configuration substantially suppresses this effect.}
\label{fig:nri_on_sky}
\end{figure}

The target for this test was a star in the NGC6025 cluster observed with an exposure time of 8 s through the Johnson $V$ filter. As shown in Figure \ref{fig:nri_on_sky}, the first amplifier measurements exhibit comparable charge levels, as indicated by the color scale. A visual inspection shows a similar shape for the star in both conditions, demonstrating that the NRI does not affect the measurement in the first amplifier, as studied in the laboratory environment in \citet{Blas_SPIE_2024}.
In contrast, the 15th amplifier measurements reveal a difference between the two configurations. The fixed-voltage configuration shows an apparent shift in the star centroid, or charge concentration distribution from left to right in the image. In the fixed-voltage case, the effect appears as a tail to the right of the center of the star, which is consistent with the NRI mechanism from the charge transfer direction. This shift is not observed when $\text{V}_{\text{ref}}$ is clocked.

Finally, the difference between the first and the 15th amplifier measurements provides a more direct visualization of the NRI effect. The images were calibrated for the clocked $\text{V}_{\text{ref}}$ condition. In the fixed $\text{V}_{\text{ref}}$ configuration, frame \ref{fig:nri_on_sky_c} shows a dipole-like residual, with a negative region on one side of the stellar profile and a positive region on the other. According to the color scale, this residual reaches $\sim10^4$ electrons in both regions for pixels that have a value close to the full well for the first amplifier. In contrast, frame \ref{fig:nri_on_sky_f}, shows no significant redistribution of the charge allocation between the 1st and 15th amplifier measurements. According to the scale, the residuals are small and are concentric to the center of the star, suggesting they are not related to transfer problems.
This on-sky result confirms that the $\text{V}_{\text{ref}}$-clocking technique effectively improves charge removal from the sense node and its reinsertion into the serial register pixels.

\subsection{Region of interest and truncated prescan mode}
\label{sec:roi}
Two non-standard readout modes were explored during the observations to investigate the capabilities of the system: a region-of-interest (ROI) mode \citep{chierchie2021smartreadout} and a truncated prescan mode \citep{Blas_SPIE_2024}. The first mode defines a region of the image to perform additional non-destructive measurements to further reduce readout noise. This increases the signal-to-noise ratio in the pre-defined region with minimal time overhead during the full sensor readout.

\begin{figure}[h!]
    \centering
  \includegraphics[width=0.8\linewidth]{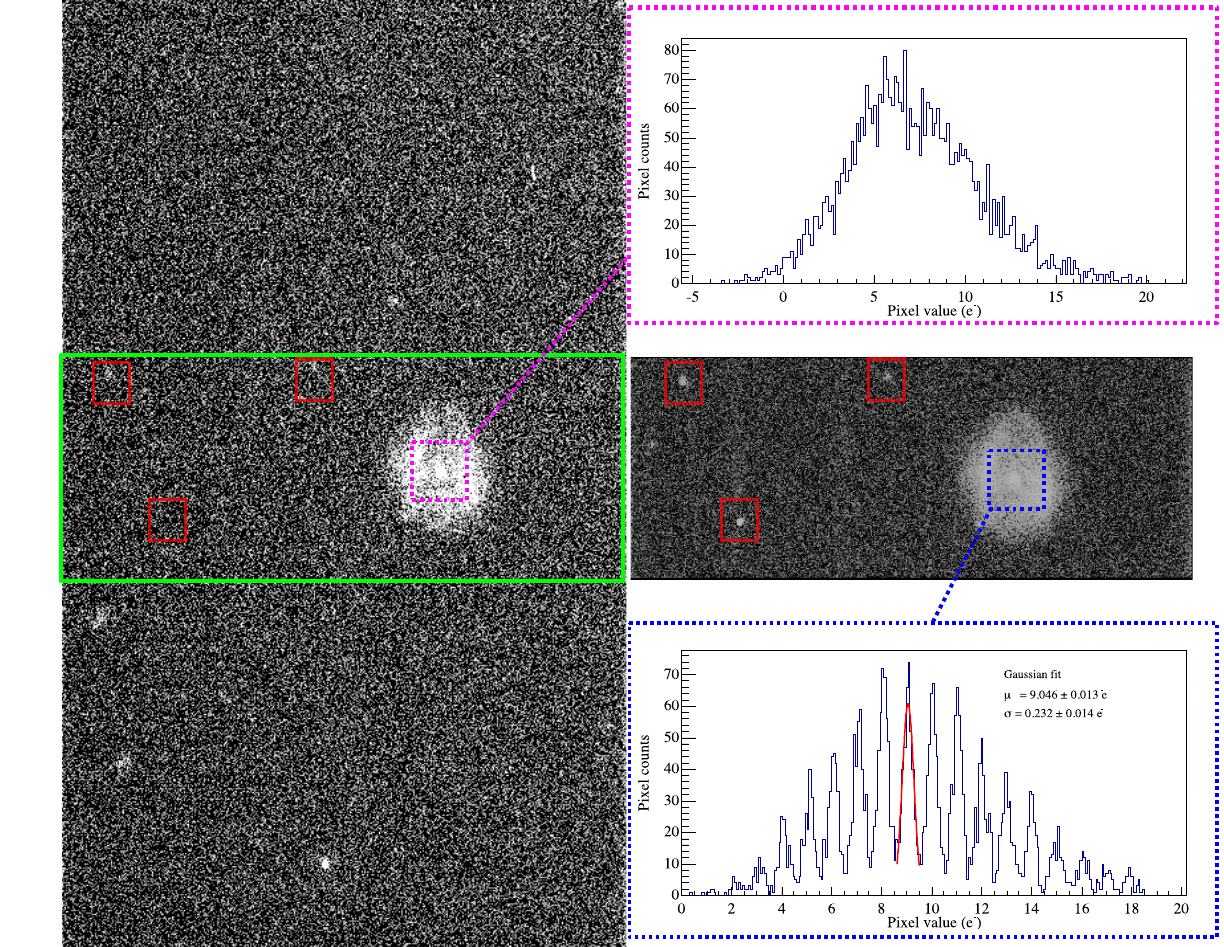}
  \caption{On the left, an image taken of the nebula NGC~6153  with single sampling ($N_S=1$). At the middle right, the region highlighted in green, has been taken with $N_S=36$ in an ROI configuration. At the top, the pixel histogram of the region framed by the magenta dashed line. At the bottom, the pixel distribution, showing single-electron resolution, is performed for the same FOV region framed in blue. Also, the image shows stars that were not visible in the normal image, which are highlighted in red.}
  \label{fig:ROI}
\end{figure}

For this purpose, the planetary nebula NGC~6153 was imaged in standard operation and ROI modes. For both cases, a 20-second exposure with the $B$ filter was used. Fig.~\ref{fig:ROI} shows representative images captured for this test. The left image corresponds to the one-sample-per-pixel readout ($N_S = 1$) of the full FOV. A region of 200 rows around the nebula was selected for the ROI (region outlined in green) using $N_S = 36$, while the rest uses $N_S = 1$. This ROI is shown in the right image portion of the figure.

The pixel distribution for the central region of the nebula is shown for the cases $N_S = 1$ and $N_S = 36$ in the top and bottom histograms, respectively. $N_S = 36$ is sufficient for this detector to achieve single-photon counting capability, as can be seen from the well-resolved peaks corresponding to quantized pixel charge from 0 to 20 e$^{-}$. The readout noise measured in each of these peaks is the same as that obtained for the full image readout with the same multiple sampling ($N_S = 36$) in Figure \ref{fig:noiseAtFermi}. The bottom histogram shows the fitted Gaussian distribution to extract its standard deviation of $0.232 \pm 0.014 $e$^{-}$.
Both images also highlight stars hidden in the noise background of the single-sample image.

The second readout mode explored, the truncated prescan mode, reduces the readout overhead associated with the extended serial register. In this mode, each row is transferred to the serial register after 512 horizontal pixel shifts of the preceding row, corresponding to the width of the active region. At this point, the 512 serial register positions beneath the active region are empty and can accommodate the subsequent row, even though the last 252 pixels of the preceding row are still propagating through the extended serial register containing the 16 amplifiers. This allows consecutive rows to be read without introducing empty pixels between them, thereby eliminating the recurring pre-scan readout overhead while preserving the measurements from all amplifiers \citep{Blas_SPIE_2024}.

\begin{figure}[h!]
\centering
\begin{subfigure}[t]{0.32\linewidth}
    \centering
    \includegraphics[trim={0 0 0 0}, clip=true, width=\linewidth]{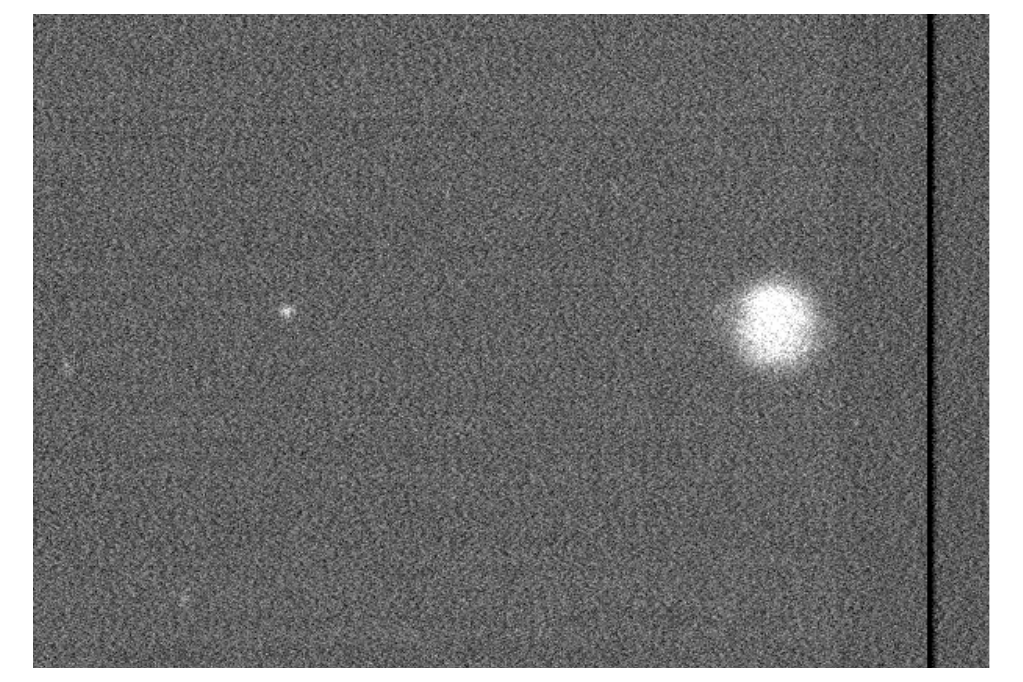}
    \caption{}
    \label{fig:continous_readout_a}
\end{subfigure}
\hfill
\begin{subfigure}[t]{0.32\linewidth}
    \centering
    \includegraphics[trim={0 0 0 0}, clip=true, width=\linewidth]{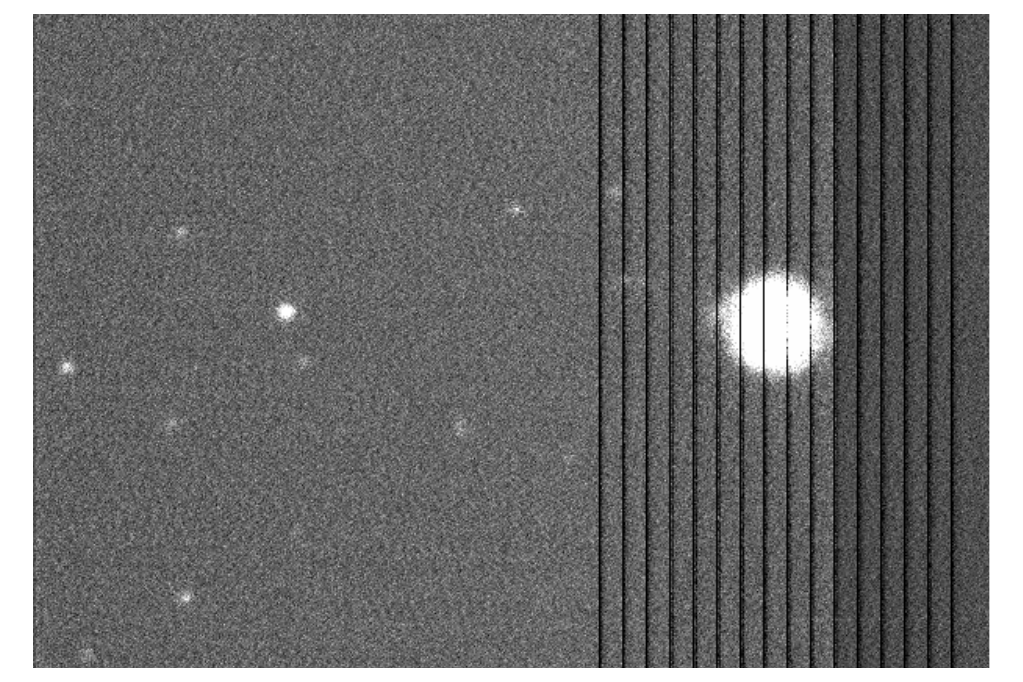}
    \caption{}
    \label{fig:continous_readout_b}
\end{subfigure}
\hfill
\begin{subfigure}[t]{0.32\linewidth}
    \centering
    \includegraphics[trim={0 0 0 0}, clip=true, width=\linewidth]{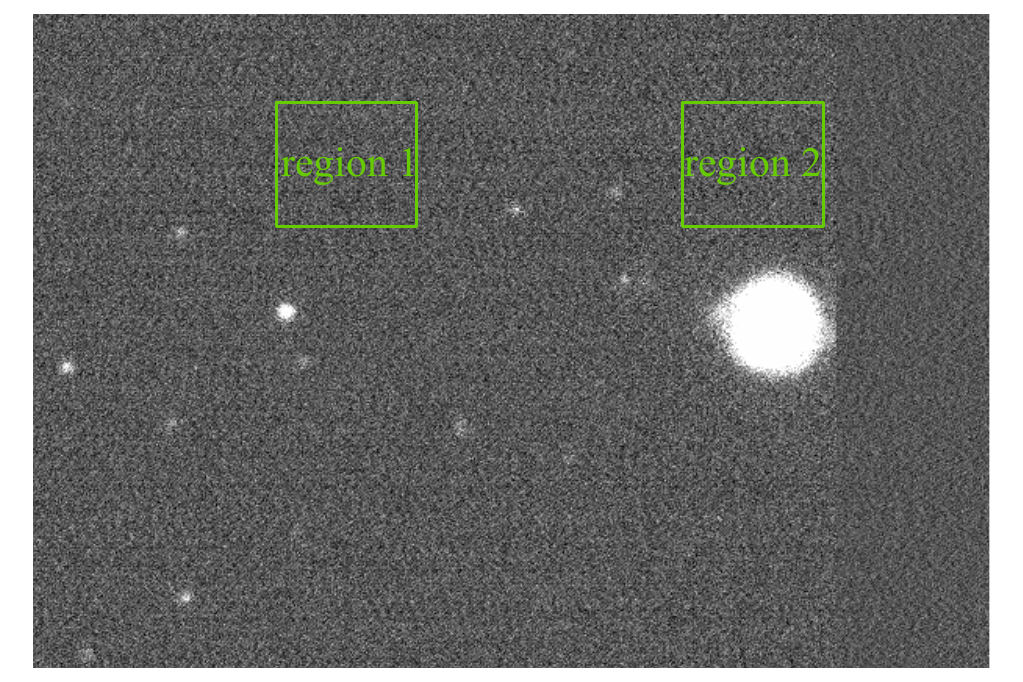}
    \caption{}
    \label{fig:continous_readout_c}
\end{subfigure}

\begin{subfigure}[t]{0.45\linewidth}
    \centering
    \includegraphics[trim={0 0 0 0}, clip=true, width=\linewidth]{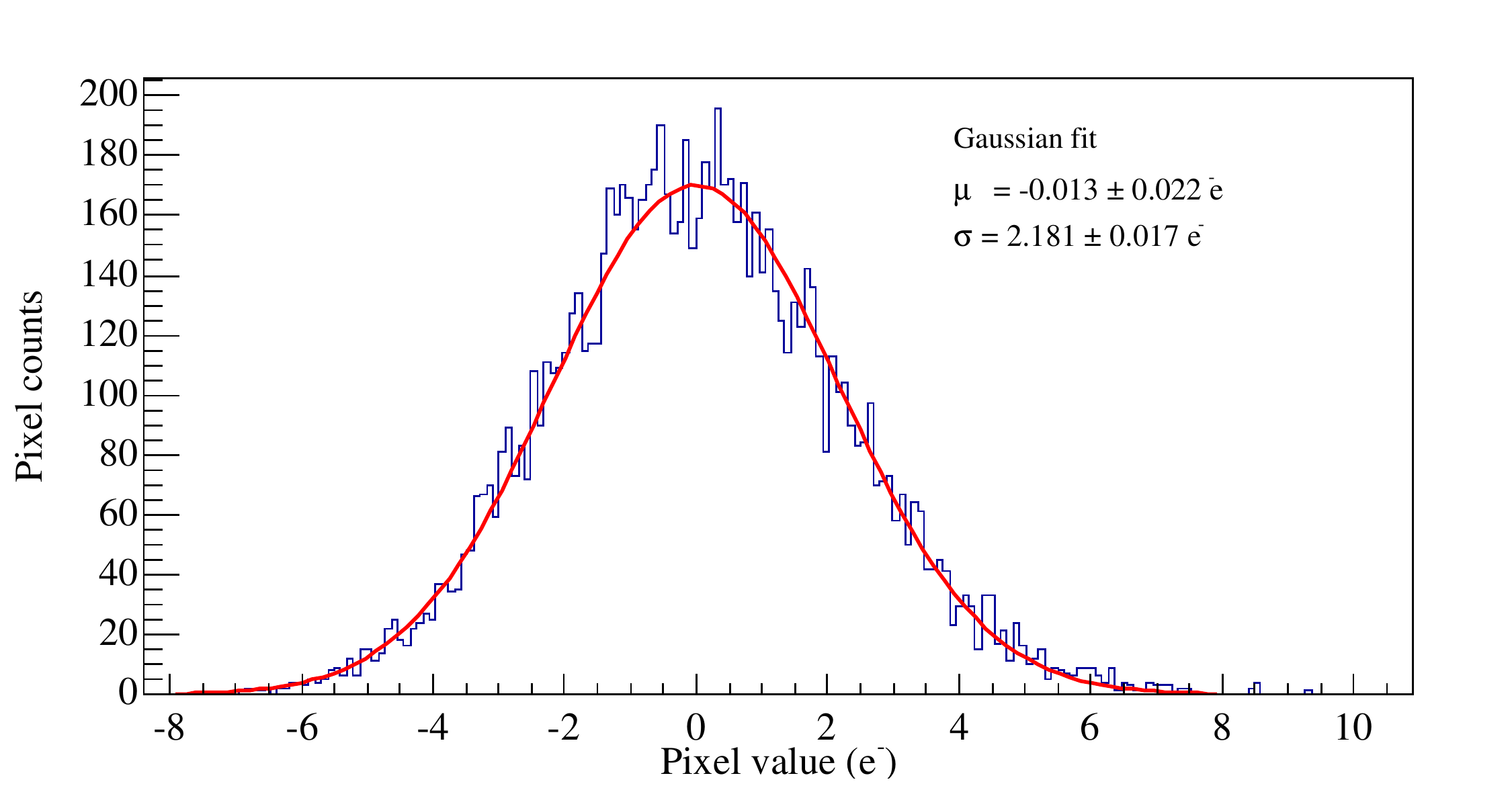}
    \caption{}
    \label{fig:continous_readout_d}
\end{subfigure}
\hfill
\begin{subfigure}[t]{0.45\linewidth}
    \centering
    \includegraphics[trim={0 0 0 0}, clip=true, width=\linewidth]{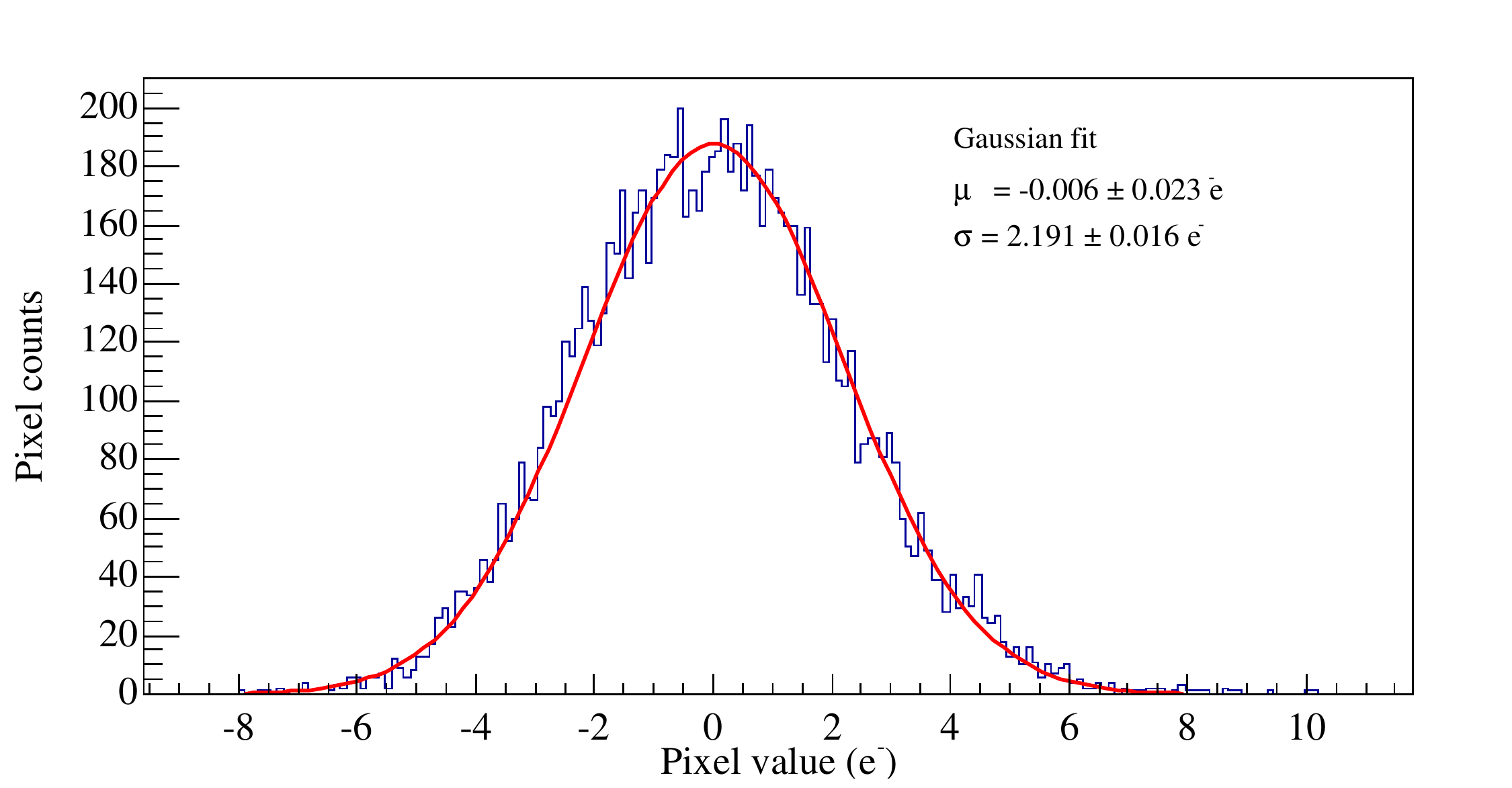}
    \caption{}
    \label{fig:continous_readout_e}
\end{subfigure}

\caption{(a) Output image of the NGC 3918 obtained from the second amplifier using the truncated prescan mode. The vertical dark line indicates baseline variation due to the vertical clocks affecting the video output signal. (b) Image obtained by combining the outputs of all 16 amplifiers.
(c) Final product image in continuous mode after subtracting the median image. Region 1 and region 2 are the selected regions in the active area with and without the effect of the baseline change by vertical clocks. (d) Histogram of pixel values from region 1 with a Gaussian fit. (e) Histogram of pixel values from region 2 with a Gaussian fit.}
\label{fig:continous_readout}

\end{figure}

A FOV around the planetary nebula NGC~3918 was used as a target in this case. Several 10-second exposures were acquired through the $B$ filter. As the row is clocked vertically before the previous row has been fully read out, they produce a change in baseline that appears in an intermediate region of the output image. Fig.\ref{fig:continous_readout_a} shows the baseline change, dark vertical region, to the right of the nebula for the second amplifier. This baseline movement affects the images from the different amplifiers at different positions. Figure \ref{fig:continous_readout_b} shows the image produced from averaging the images from the different channels. The first step was to evaluate the stability of the baseline in order to remove it. For this, 15 bias images, with shutter closed and 0 s exposure time, acquired in this readout mode were combined into a median image, which was then used to correct the baseline variation in the combined image.
The corrected image is shown in Fig.~\ref{fig:continous_readout_c}. To assess whether this correction affected the readout noise, histograms were produced from two regions: one separated from the baseline changes and one with pixels overlapping them. The resulting histograms and their corresponding Gaussian fits are shown in Fig.~\ref{fig:continous_readout_d} and ~\ref{fig:continous_readout_e}, respectively. The fitted standard deviation and errors for both distributions are $2.181\pm 0.017$e$^-$ and $2.191\pm 0.016$e$^-$, respectively, showing good statistical agreement between the two regions.

For our sensor, this readout scheme eliminates the readout of 288 columns, reducing the total image readout time by 5.3 s ($\sim31.72\%$) while preserving the same readout-noise performance.

\subsection{Low-Noise Open-Shutter Continuous Readout Mode}
\label{sec:continuous_readout}

\begin{figure}[t]
\centering

\begin{subfigure}[c]{0.32\linewidth}
    \centering
    \includegraphics[width=\linewidth]{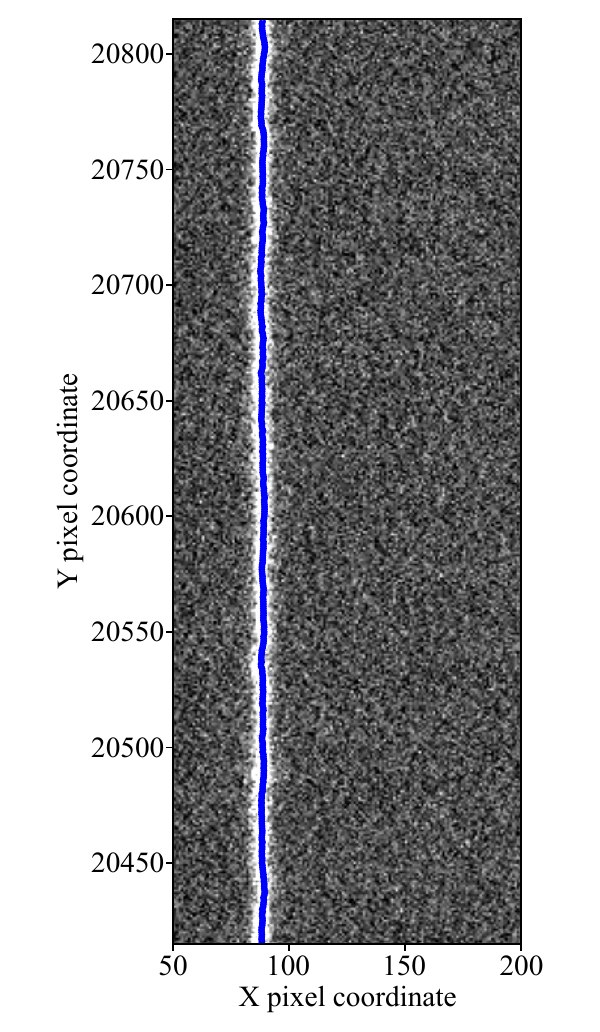}
    \caption{}
    \label{fig:continuous_readout_a}
\end{subfigure}
\hfill
\begin{minipage}[c]{0.65\linewidth}
    \centering
    \begin{subfigure}[c]{\linewidth}
        \centering
        \includegraphics[width=0.81\linewidth]{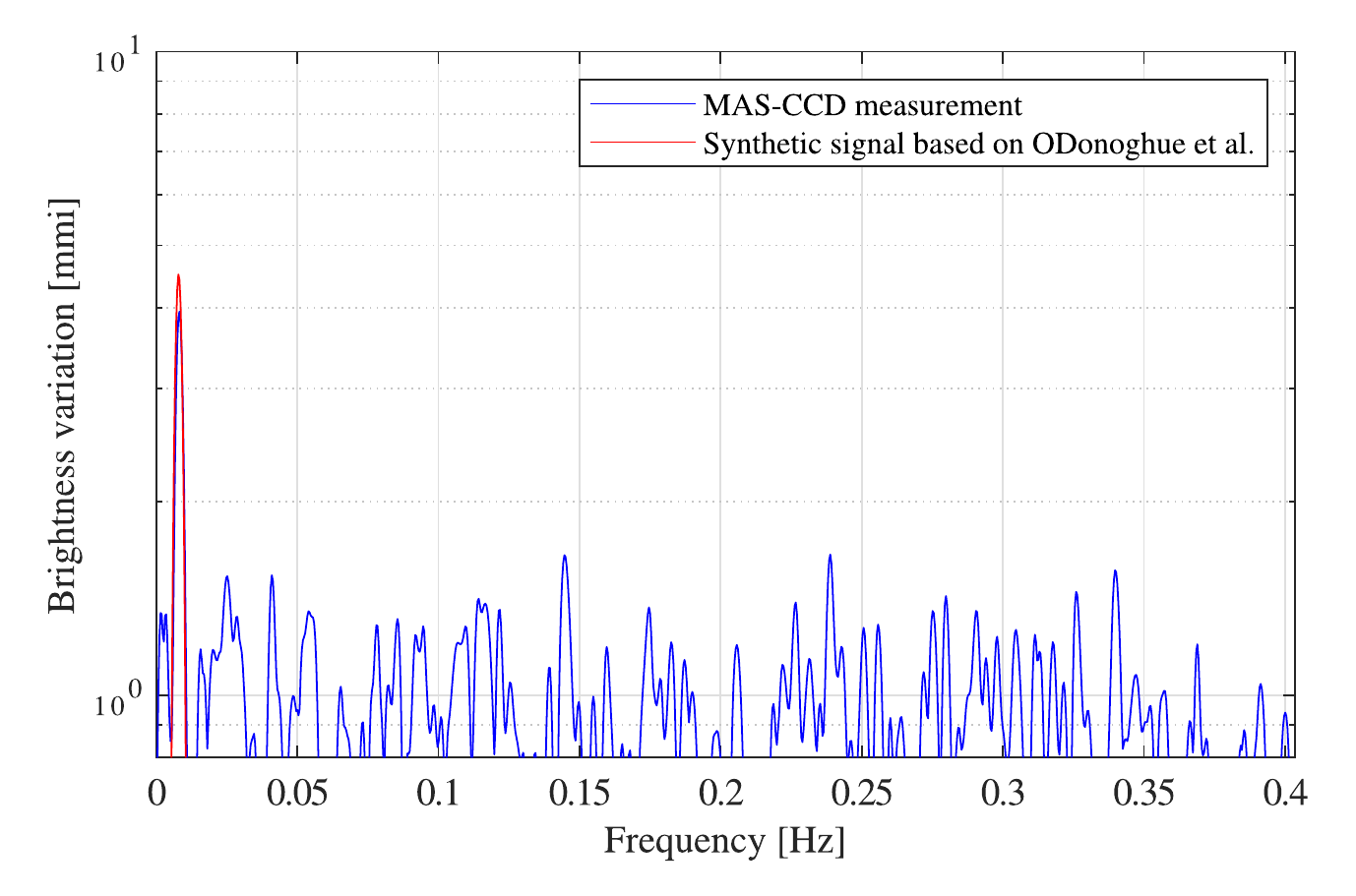}
        \caption{}
        \label{fig:continuous_readout_b}
    \end{subfigure}

    \vspace{4pt}

    \begin{subfigure}[c]{\linewidth}
        \centering
        \includegraphics[width=0.87\linewidth]{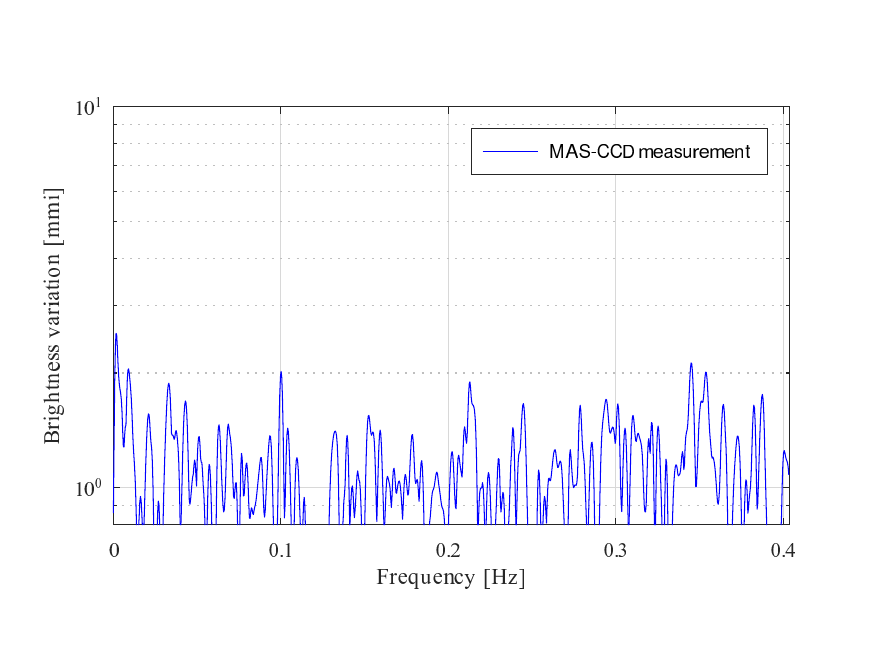}
        \caption{}
        \label{fig:continuous_readout_c}
    \end{subfigure}
\end{minipage}
\caption{Open shutter continuous readout observations with the MAS-CCD. (a) Fragment of the PB~8783 star trail (white), with the extracted centroid shown as a blue line. Time increases along the vertical axis. (b) Fourier amplitude spectrum of PB~8783 averaged over three observations, compared with a synthetic signal generated by summing in-phase the dominant pulsation modes reported by \cite{ODonoghue1998}. The prominent peak near 8.3 mHz is consistent with the known stellar oscillations. (c) Amplitude spectrum of HD~37836 averaged over three observations, showing no comparable prominent periodic feature.}
\label{fig:continuous_readout}
\end{figure}

Another non-standard operational mode is open-shutter continuous readout (e.g., \cite{Bianco_2009}). In this mode, the shutter is left open and the detector is clocked continuously, producing a trail of charge for each bright object in the field of view (Figure~\ref{fig:continuous_readout_a}). The CCD image thus becomes a flux time series sampled at the row readout time. The analysis of continuous readout data is complicated by the atmospheric seeing, telescope vibration, and mechanical instabilities, which correlate the signal in nearby rows. While a detailed analysis of continuous readout data is beyond the scope of this paper, a recent example is presented in \cite{daniels2026}.

Observations with the MAS-CCD were conducted in continuous readout mode, using the low-pixel-signal-voltage configuration listed in Table~\ref{tab:voltages}, to collect 50,000 rows per observation with the shutter open. The effective sampling time is given by the full row readout time, $T_s = 12.39$~ms ($\sim$80~Hz). The first 1,024 rows, which were in the active region when the readout started and therefore have non-uniform exposure times, were discarded; the remaining 48,976 rows correspond to a total acquisition time of approximately 10.1~minutes per observation. Two bright stars were observed: the rapidly pulsating hot subdwarf PB~8783 ($B \approx 12.2$~mag) \citep{Hog2000} and the luminous blue supergiant HD~37836 in the Large Magellanic Cloud ($B \approx 10.7$~mag) \citep{Zacharias_2013}. These data were collected to demonstrate the viability of the continuous readout observing mode and the stability of the system.

A set of 15 $B$-band continuous readout observations was collected for PB~8783. For each observation, a photometric time series was assembled by locating the star trail, performing background correction, and extracting the centroid and magnitude for each row. The time series were median filtered with a 65-row window to remove outliers caused by cosmic rays, and then down-sampled by a factor of 100 to reduce the data volume; the resulting sampling time of 1.24 s is much shorter than the $\sim$120~s pulsation period of PB 8783 and the total observation time of ~600s covers only a few periods. The stellar intensity of each observation was normalized by its median value to correct for scaling variations between observations. Three of the 15 observations were selected for spectral analysis based on the absence of noticeable contamination from atmospheric conditions, such as cloud coverage, or other observational disturbances. Each selected time series was windowed with a Hann function, its spectrum was computed with a zero-padded fast Fourier transform, and the spectra of the three observations were averaged.

As a reference, a synthetic signal was generated by summing in phase the 11 dominant pulsation modes reported by \cite{ODonoghue1998}, which span 7.3--10.6~mHz. Amplitudes are expressed in milli-modulation intensity (mmi), the linear unit adopted by \cite{ODonoghue1998}, in which 10~mmi corresponds to a 1\% change in brightness. The synthetic signal was processed in the same way as the measurements. Because each observation is short, the frequency resolution, $1/T \approx 1.6$~mHz, is not sufficient to distinguish the individual modes, most of which lie within 1~mHz of each other. Moreover, the relative phases of the modes at the time of our observations are unknown, and the in-phase sum represents one possible realization of their combined signal. The synthetic spectrum therefore serves as a reference for the expected location and level of the pulsation signal, not for the amplitudes or frequencies of individual modes.

The amplitude spectrum of PB~8783 measured with the MAS-CCD (Figure~\ref{fig:continuous_readout_b}) shows a peak at $\sim$8.3~mHz with an amplitude of 3.94~mmi, above a noise level of 1~mmi (S/N~$=$~4). The position and amplitude of this peak are consistent with those of the synthetic signal. The peak is a blend of unresolved modes, and \cite{ODonoghue1998} report variations in the amplitude, of the order of 3~mmi, and in the phase of the pulsations over time, so some differences between the measurement and the synthetic signal are expected.

The same procedure was applied to 16 $B$-band observations of the luminous supergiant HD~37836, of which three were selected with the same criterion. The known photometric variability of HD~37836 occurs on timescales much longer than the $\sim$10-minute duration of our observations \citep{Zacharias_2013}, so no coherent signal is expected in the mHz band. Its spectrum (Figure~\ref{fig:continuous_readout_c}) shows no comparable periodic feature.

These results show that the continuous readout mode of the MAS-CCD recovers a known stellar oscillation in the mHz band, while no comparable periodic signal is detected in observations of HD~37836.

\subsection{RGB Composition}

The observation run included multi-band imaging of targets using the Johnson-Cousins red ($R$), green ($V$), and blue ($B$) photometric filters. For each filter, three images were acquired at $N_S = 1$ with exposure times of 50~s (red), 100~s (green), and 200~s (blue). Each channel was optimally weighted to achieve the lowest noise in the final image for each filter. These were later reduced by applying bias, dark, and flat-field corrections. Calibration data was taken using dome flat lights after the run, in the early morning.

\begin{figure}[h!]
\centering
\begin{subfigure}[t]{0.25\linewidth}
    \centering
    \includegraphics[trim={0 0 0 0}, clip=true, width=\linewidth]{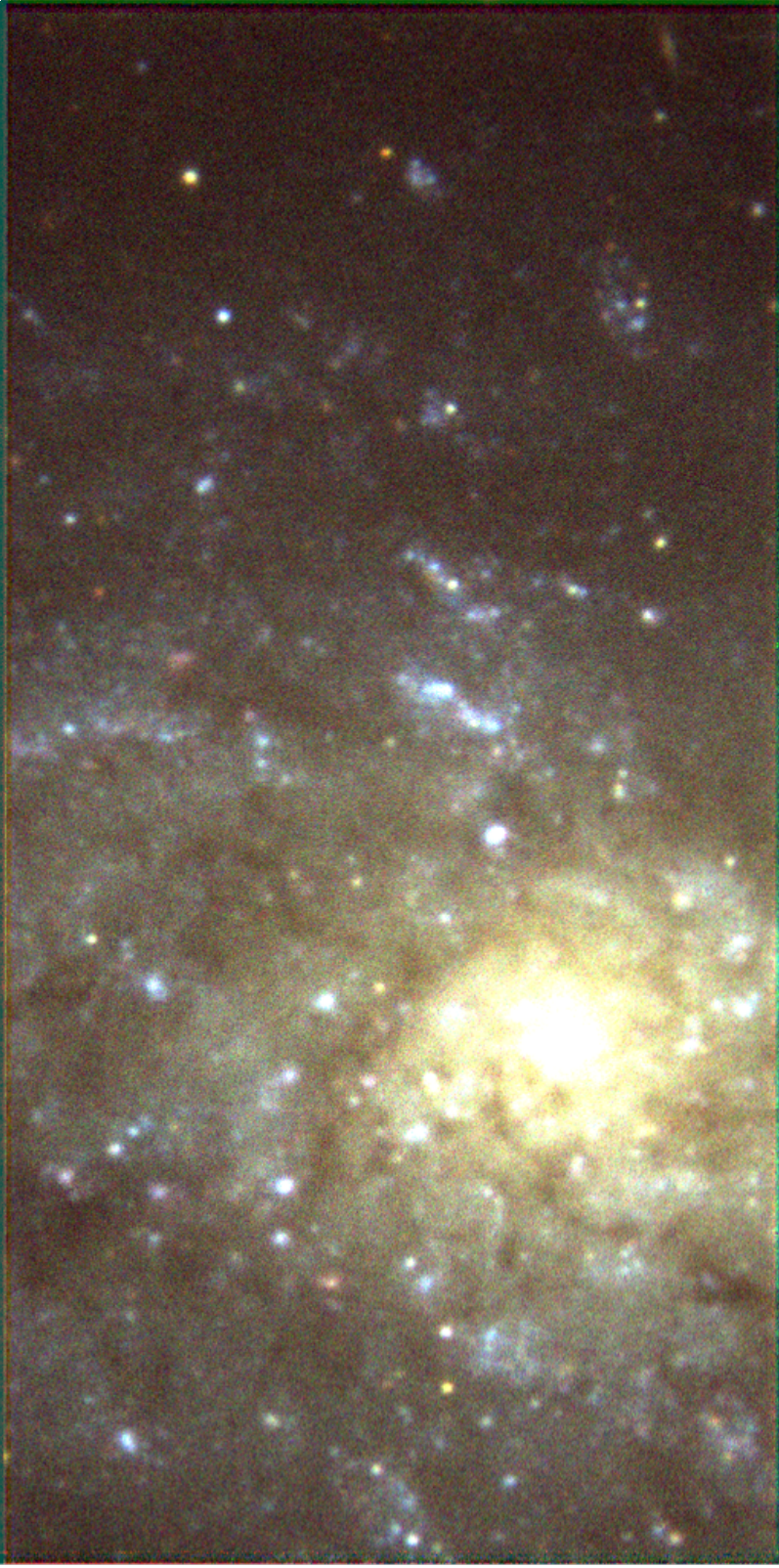}
    \caption{}
    \label{fig:astro-photos_a}
\end{subfigure}
\begin{subfigure}[t]{0.25\linewidth}
    \centering
    \includegraphics[trim={0 0 0 0}, clip=true, width=\linewidth]{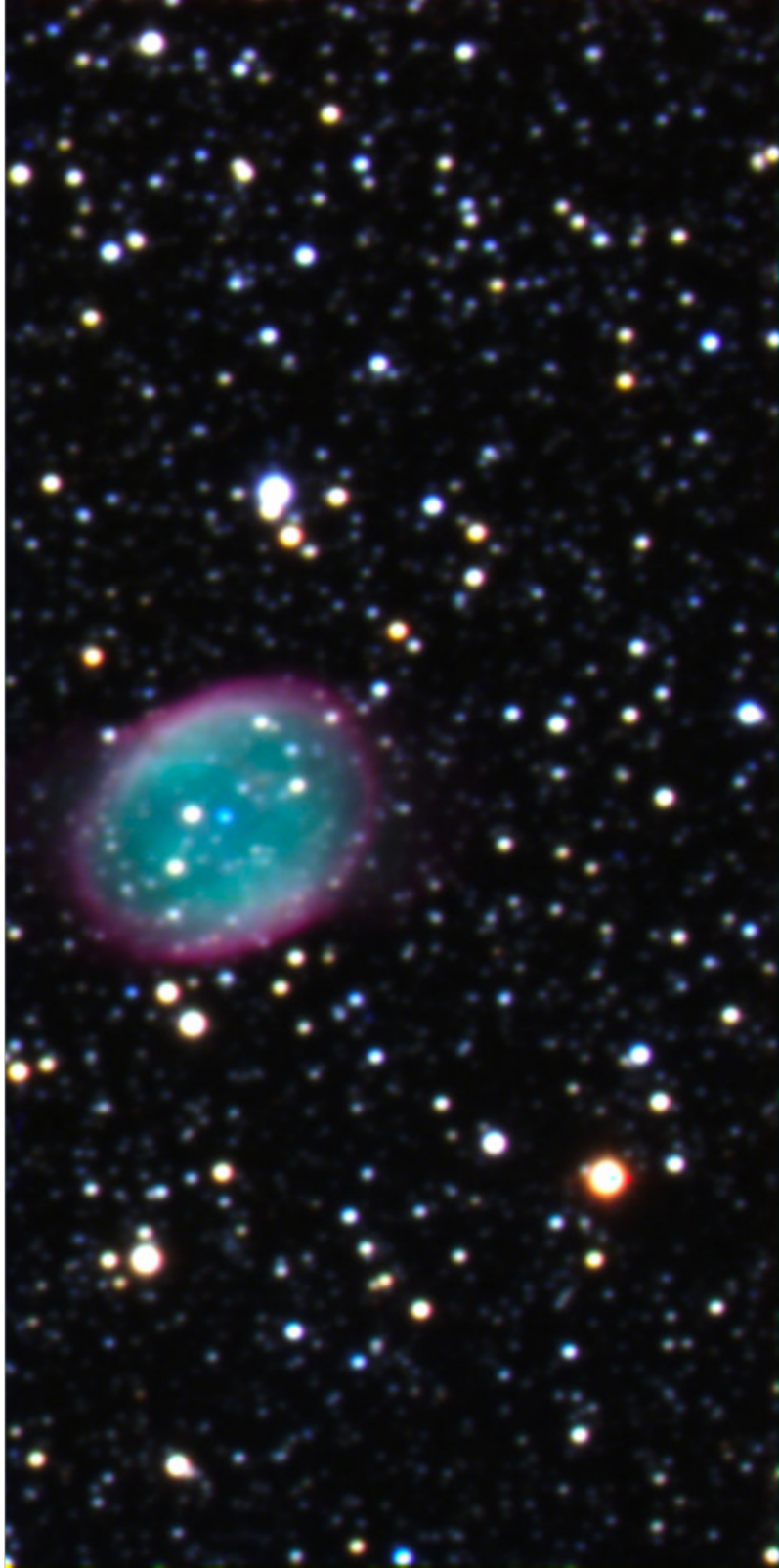}
    \caption{}
    \label{fig:astro-photos_b}
\end{subfigure}

\caption{Multi-band imaging demonstration for targets (a) NGC 7793 and (b) NGC 6563.}
\label{fig:astro-photos}

\end{figure}

Image registration was performed using custom designed software tools in two stages: first on a per-filter basis prior to constructing master frames for each filter to remove the cosmic rays, and then before combining the different wavelength channels. Figure~\ref{fig:astro-photos} shows false-color RGB images for two targets, NGC~7793 and NGC~6563.

It is important to note that, due to the multiple-output-stage architecture, 16 images are produced with independent noise realizations for each color band. Therefore, it was necessary to combine a total of 48 products for each RGB composite image.

\section{Conclusions} \label{sec:conclusions}
The results presented in this work constitute the first on-sky demonstration of MAS-CCD technology. The observation run enabled the validation of techniques previously studied only under laboratory conditions. These techniques were found to perform as expected when deployed on-sky.

The NRI mitigation method proved effective in improving the performance of the device for large charge packets, which would have otherwise been limited by transfer inefficiencies. Two novel readout schemes were demonstrated to reduce readout times while preserving the key advantages of the MAS-CCD. In particular, the ROI strategy enabled high-quality imaging at the single-photon counting level while minimizing the impact of multiple sampling on the total readout time.

A science case study was also performed, consisting of the analysis of a known variable star using a continuous-readout scheme with the shutter open. This approach enabled the acquisition of a temporal trace at a defined sampling frequency given by the sensor's readout sequence ($T_s=12.39$ ms). Observations of the variable star PB 8783 show a prominent peak near 8.3 mHz consistent with the literature. This demonstrates the capability and potential of the MAS-CCD technology.

Finally, except for the higher readout noise observed at the telescope, the overall performance of the device was found to be consistent with that achieved at Fermilab. The noise analysis framework developed in this work provides a tool to identify performance limitations and guide further optimization in future experiments.

\begin{acknowledgments}

The MAS-CCD was developed as a collaborative endeavor between Lawrence Berkeley National Laboratory and Fermi National Accelerator Laboratory. Funding for the design and fabrication of the MAS device described in this work came from a combination of sources, including the DOE Quantum Information Science (QIS) initiative, the DOE Early Career Research Program, and the Laboratory Directed Research and Development Program at Fermi National Accelerator Laboratory under Contract No. DE-AC02-07CH11359. This research has been partially supported by the Heising-Simons Foundation (Grant No. 2023-4611), Guillermo Fernandez Moroni’s DOE Early Career research program, and by NASA APRA award No.\ 80NSSC22K1411.

Funding for the design, fabrication, and testing of part of the mechanics and electronics, as well as the on-site laboratory infrastructure used for this project, was provided by NSF’s NOIRLab, which is managed by the Association of Universities for Research in Astronomy (AURA) under a cooperative agreement with the National Science Foundation.

Christopher Montalban acknowledges the financial support of DIDULS/ULS, through the project PTE2453858.

This research has used data acquired at the CTIO/SMARTS 0.9m telescope, which is operated as part of the SMARTS Observatory by RECONS (www.recons.org) members Todd Henry, Wei-Chun Jao, and Aman Kar.

\end{acknowledgments}

\facilities{Fermi National Accelerator Laboratory (CCD Lab), Recinto Aura, Cerro-Tololo Interamerican Observatory 0.9m SMARTS Telescope}.

\software{Astropy, SExtractor, ROOT}.

\bibliography{main}{}
\bibliographystyle{aasjournalv7}

\end{document}